\documentclass[12pt]{article}

\usepackage{lineno,hyperref}
\modulolinenumbers[5]

\usepackage{graphicx}
\usepackage{newtxtext}
\usepackage{hyperref}
\hypersetup{
    colorlinks = true,
    urlcolor   = blue,
    citecolor  = black,
}

\newcommand{\RomanNumeralCaps}[1]
\linenumbers

\usepackage{amsmath,amssymb,esint,bbm,textgreek,mathrsfs,mathtools}
\usepackage{setspace}
\usepackage{etoolbox}

\usepackage{stackengine}
\usepackage{scalerel}

\usepackage{cleveref}
\usepackage{algpseudocode}
\usepackage{algorithm}
\usepackage{caption}
\usepackage{subfigure}
\usepackage[normalem]{ulem}
\usepackage[table]{xcolor}
\usepackage{booktabs}
\usepackage{placeins}

\usepackage{cleveref}
\crefformat{equation}{(#2#1#3)}
\usepackage{algpseudocode}
\usepackage{algorithm}
\usepackage{caption}
\usepackage{subfigure}
\usepackage{graphicx}
\usepackage[normalem]{ulem}
\usepackage{color, colortbl}
\usepackage{booktabs}
\usepackage{pdflscape}
\graphicspath{{Figs/}}
\usepackage{Matharu_commands}

\begin{document}

\title{On Maximum Enstrophy Dissipation in $2$D Navier-Stokes Flows in
  the Limit of Vanishing Viscosity}



\author{Pritpal Matharu$^{1,}$\thanks{International Research Fellow of
    the Japan Society for the Promotion of Science (Graduate
    Fellowships for Research in Japan)},  \ Tsuyoshi Yoneda$^2$ and
  Bartosz Protas$^{1,}$\thanks{Email address for correspondence: {\tt
      bprotas@mcmaster.ca}} \\ 
	$^1$Department of Mathematics and Statistics, McMaster University \\
	Hamilton, Ontario, L8S 4K1, Canada \\
        $^2$Graduate School of Economics, Hitotsubashi University \\ 2-1 Naka, Kunitachi, Tokyo 186-8601, Japan
      }
\date{\today}

\maketitle

\begin{abstract}
  We consider enstrophy dissipation in two-dimensional (2D)
  Navier-Stokes flows and focus on how this quantity behaves in the
  limit of vanishing viscosity. After recalling a number of {a
    priori} estimates providing lower and upper bounds on this
  quantity, we state an optimization problem aimed at probing the
  sharpness of these estimates as functions of viscosity. More
  precisely, solutions of this problem are the initial conditions with
  fixed palinstrophy and possessing the property that the resulting 2D
  Navier-Stokes flows locally maximize the enstrophy dissipation over
  a given time window.  This problem is solved numerically with an
  adjoint-based gradient ascent method and solutions obtained for a
  broad range of viscosities and lengths of the time window reveal the
  presence of multiple branches of local maximizers, each associated
  with a distinct mechanism for the amplification of palinstrophy. The
  dependence of the maximum enstrophy dissipation on viscosity is
  {shown to be} in quantitative agreement with the estimate due
  to Ciampa, Crippa \& Spirito (2021), demonstrating the sharpness of
  this bound.
\end{abstract}

{\bf{Keywords:}}   2D Navier-Stokes equation; enstrophy dissipation; inviscid
  limit; PDE-constrained optimization

\section{Introduction}
\label{sec:intro}

The physical phenomenon of ``anomalous dissipation'', also referred to
as the ``zeroth law of turbulence'', is one of the oldest problems in
turbulence \cite{frisch1995turbulence}. This empirical law states that
the energy dissipation in either forced or decaying three-dimensional
(3D) turbulent flows approaches a nonzero limit as the fluid viscosity
$\nu > 0$ vanishes, all other flow parameters remaining fixed. There
is a lot of evidence coming from both experiments and numerical
simulations supporting this anomalous behavior of the energy
dissipation \cite{Sreenivasan1998,YeungZhaiSreenivasan2015}, but we
are still far from being able to understand this problem from the
mathematical point of view. The main consequence of the dissipation
anomaly is an unbounded increase of velocity gradients which would in
turn imply finite-time singularities in solutions of the inviscid
Euler equations \cite{Dascaliuc2012}. Similar dissipation anomalies
are also known to occur in the behavior of passive scalars
\cite{Sreenivasan2019,Mazzucato2022}.

Dissipation anomaly arises in solutions of the one-dimensional (1D)
Burgers equation \cite{EyinkDrivas2015}. As regards 2D flows, the
relevant question is about the behavior of the enstrophy dissipation
in the limit of vanishing viscosity. The assumption that enstrophy
dissipation tends to a finite (nonzero) limit as $\nu \rightarrow 0$
underlaid Batchelor's theory of 2D turbulence
\cite{Batchelor1969}. However, in \cite{Tran2006} it was argued that
this quantity in fact vanishes in the inviscid limit such that
Navier-Stokes flows in 2D are not subject to dissipation anomaly.
This result was confirmed by rigorous analysis of the inviscid limit
of 2D Navier-Stokes flows \cite{Lopes2006}.

While there is no dissipation anomaly in 2D flows, it is interesting
to know the worst-case (slowest) rate at which the enstrophy
dissipation vanishes in the limit $\nu \rightarrow 0$. A number of
theoretical results, in the form of both lower and upper bounds on the
dependence of the enstrophy dissipation on $\nu$, have been
established and are reviewed below. The goal of the present study is
to address this question computationally by finding flows with the
largest possible enstrophy dissipation as the viscosity vanishes. Such
``extreme'' flows will be found by solving suitably defined
optimization problems with constraints in the form of partial
differential equations (PDEs).  This will provide insights about the
sharpness of various rigorous bounds on the enstrophy dissipation in
the inviscid limit. While methods of PDE optimization have had a long
history in various applied areas \cite{g03}, they have recently been
employed to study certain fundamental problems concerning extreme
behavior in fluid mechanics \cite{p21a}. In particular, problems
somewhat related to the subject of the present study were investigated
using such techniques in \cite{ap13a,ap13b,ayala_doering_simon_2018}.

We consider the incompressible Navier-Stokes system on a 2D periodic
domain $\Omega := \TT^2 = [0, 1]^2$ (``$:=$'' means ``equal to by
definition'') which can be written in the vorticity form as
\begin{subequations} \label{eq:2DNS}
\begin{alignat}{2} 
\p{\wn}{t} + \bfgrad^{\perp} \psin \cdot \bfgrad \wn &= \nu \bfDelta \wn &  \qquad &\text{in} \ \Omega \times (0, T],  \label{eq:vort_eqn} \\
-\bfDelta \psin &= \wn & \qquad &\text{in} \ \Omega \times (0, T], \label{eq:stream_vort} \\
\wn(t=0) &= \varphi &  &\text{in} \ \Omega,  \label{eq:IC} 
\end{alignat} 
\end{subequations}
where $\wn$ and $\psin$ are the vorticity component perpendicular to
the plane of motion and the corresponding streamfunction, both assumed
to satisfy the periodic boundary conditions in the space variable
$\x$, whereas $T > 0$ is the length of the time window considered. The
symbol $\varphi$ denotes the initial condition which without loss of
generality is assumed to have zero mean, i.e., 
\begin{equation}
\intO \varphi(\x) \, d\x = 0. \label{eq:ICmean}
\end{equation}
Problem \eqref{eq:2DNS} is known to be globally well-posed in the
classical sense \cite{kl04}. Its solutions are characterized by the
enstrophy and palinstrophy defined, respectively, as\footnote{For
  consistency with the convention used in our earlier studies,
  cf.~\cite{p21a}, both these quantities are defined with a factor of
  1/2. Without the risk of confusion we will sometimes use the
  simplified notation $\E(t) = \E(\wn(\cdot, t))$ and $\P(t) =
  \P(\wn(\cdot, t))$.}
\begin{align}
\E(\wn(\cdot, t)) &:= \frac{1}{2} \, \intO \left| \wn(\x, t) \right|^2 \, d\x, \label{eq:E}\\
\P(\wn(\cdot, t)) &:= \frac{1}{2} \, \intO \left| \bfgrad\wn(\x, t) \right|^2 \, d\x, \label{eq:P}
\end{align}
which satisfy the relation
\begin{equation} \label{eq:dEdt}
\frac{d\E(t)}{dt} =  - 2\nu\P(t).
\end{equation}
We then define our main quantity of interest as 
\begin{equation} 
\label{eq:chi}
\chin(\varphi) := \frac{2 \nu}{T} \int_0^T \P(t)\, dt 
=  {\frac{\nu}{T}} \, \int_{0}^{T} \intO \left| \bfgrad\wn(\x, t; \varphi) \right|^2 \, d\x dt 
= \frac{\E(0) - \E(T)}{T},
\end{equation} 
which represents the enstrophy dissipation per unit of time and will
be viewed here as a function of the initial data $\varphi$.

The enstrophy dissipation \eqref{eq:chi} has been the subject of
numerous estimates. We refer to the following result as a
``conjecture'' since it relies on some assumptions, albeit well
justified, about the form of the spectrum of the solutions of
\eqref{eq:2DNS}.
\begin{conjecture}[Tran \& Dritschel \cite{Tran2006}] 
\label{conj:TD}
The enstrophy dissipation in solutions of system \cref{eq:2DNS} is
bounded above by
\begin{equation} 
\label{eq:Tran}
\chin \leq C\,\left[-\ln(\nu)\right]^{-\frac{1}{2}},
\end{equation}
for some constant $C>0$ depending on the initial condition $\varphi$
and the length $T$ of the time window.
\end{conjecture}
Hereafter {$C = C(T)$ will denote a generic positive constant
  depending on the length $T$ of the considered time window} with
numerical values differing from one instant to another.

Bounds on enstrophy dissipation are closely related to another problem
which has recently received considerable attention, namely, the
question of the convergence as $\nu \rightarrow 0$ of Navier-Stokes
flows to solutions of the inviscid Euler equations obtained by setting
$\nu = 0$ in \eqref{eq:vort_eqn} and corresponding to the same initial
condition $\varphi$. More specifically, noting \eqref{eq:dEdt}, the
fact that solutions of the inviscid Euler system conserve the
enstrophy and using the reverse triangle inequality, we have
\begin{align}
\chin(\varphi)  &= {\frac{\nu}{T}} \, \int_{0}^{T} \left\| \bfgrad\wn(\x, t; \varphi) \right\|^2_{L^2(\Omega)} \, dt = \frac{2 \nu}{T} \, \int_{0}^{T} \P(t) \, dt  \nonumber \\
&= \frac{1}{T} \, \left[\E(0) - \E(T)\right] = \frac{1}{T} \, \left[ \left\| \varphi \right\|^2_{L^2(\Omega)} - \left\| \wn(\x, T; \varphi) \right\|^2_{L^2(\Omega)} \right] \nonumber \\
&= \frac{1}{T} \, \left[ \left\| \w(\x, T; \varphi) \right\|^2_{L^2(\Omega)} - \left\| \wn(\x, T; \varphi) \right\|^2_{L^2(\Omega)} \right] \nonumber \\
&\leq \frac{1}{T} \, \left[ \left\| \w(\x, T; \varphi) \right\|_{L^2(\Omega)} + \left\| \wn(\x, T; \varphi) \right\|_{L^2(\Omega)} \right] \, \left\| \w(\x, T; \varphi) - \wn(\x, T; \varphi) \right\|_{L^2(\Omega)} \nonumber \\
&\leq \frac{2}{T} \, \left\| \varphi \right\|_{L^2(\Omega)}  \, \left\| \w(\x, T; \varphi) - \wn(\x, T; \varphi) \right\|_{L^2(\Omega)}, \label{eq:chibound}
\end{align} 
where $\omega(\x,t) := \omega_0(\x,t)$ denotes the vorticity in the
inviscid Euler flow. The above relation shows that the enstrophy
dissipation over the time window $[0,T]$ can be bounded from above in
terms of the difference of the vorticity fields in the viscous and
inviscid flows obtained with the same initial data $\varphi$ at time
$t = T$.  Quantifying this difference in terms of viscosity as $\nu
\rightarrow 0$ has been the subject of some recent studies. In
\cite{ConstantinDrivasElgindi2022} the authors showed the strong
convergence of $\wn$ to $\w$ as $\nu \rightarrow 0$ when $\varphi \in
L^{\infty}(\Omega)$, implying the vanishing of the right-hand side
(RHS) in \eqref{eq:chibound}. Moreover, the following estimate was
established in the case when $\varphi \in L^{\infty}(\Omega) \cap
B_{2,\infty}^s(\Omega)$ for some $s>0$, where $L^p$ and
$B^s_{p,q}$ are the usual Lebesgue and Besov spaces, 
\begin{equation}
\sup_{t \in [0,T]} \left\| \w(\cdot, t) - \wn(\cdot,t)
\right\|_{L^p(\Omega)} \le {C}  (\nu T)^{\frac{s\,e^{-2CTM}}{p(1+s\,e^{-CTM})}},
\label{eq:chiCDE}
\end{equation}
where $M := \|\varphi\|_{L^\infty(\Omega)}$. This problem was revisited in \cite{Ciampa2021}
where it was proved that
\begin{equation}
\sup_{t \in [0,T]} \left\| \w(\cdot, t) - \wn(\cdot,t) \right\|_{L^p(\Omega)} \le C\, M^{1-\frac{1}{p}}\max\left\{\phi_{\varphi, p, M}(C\, \nu^{\frac{e^{-CT}}{2}}), \left(C\, \nu^{\frac{e^{-CT}}{2}}\right)^{\frac{e^{-CT}}{2p}} \right\}, 
\label{eq:chiCCS}
\end{equation}
where now $C = C(T, M)$ and $\phi_{\varphi, p, M} \: : \: \RR^+ \to
\RR^+$ is a continuous function such that $\phi_{\varphi, p, M}(0)=0$.
Additional results were also obtained recently in
\cite{NussenzveigLopes2021,Seis2021}. In particular, the following
bound was produced in \cite{Seis2021}, which improves the rate of
the weak convergence of $\wn$ to $\w$ as $\nu \rightarrow 0$,
\begin{equation}
\sup_{t \in [0,T]} \left\| \w(\cdot, t) - \wn(\cdot,t)
\right\|_{\dot{H}^{-1}(\Omega)} \le C \,
\left[\frac{\nu}{|\ln(\nu)|}\right]^{\frac{e^{-CT}}{2}}.
\label{eq:chiS}
\end{equation}
We reiterate that, in the light of relation \eqref{eq:chibound},
inequalities {\eqref{eq:chiCDE}--\eqref{eq:chiCCS} imply
  viscosity-dependent upper bounds on the enstrophy dissipation
  \eqref{eq:chi}. This is not the case for estimate \eqref{eq:chiS} as
  it involves a weaker norm than in \eqref{eq:chibound}. We will
  nonetheless refer to this estimate when we discuss our results in
  Section \ref{sec:results} with the hope that our findings may
  inspire further work on refining this estimate.}  On the other
hand, as is evident from the following theorem, a lower bound on the
maximum enstrophy dissipation is also available.
\begin{theorem}[Jeong \& Yoneda \cite{Jeong2021}] 
\label{thm:lower}
Let $\wn$ be the unique solution to \cref{eq:2DNS}. Then, there exists
initial data $\varphi$  such that the enstrophy
dissipation is bounded below by
\begin{equation} \label{eq:chiJY}
\chin \ge {C} \nu \, \left[-\ln(\nu)\right]^{\frac{1}{2}}.
\end{equation}
\end{theorem}

Upper bounds on the energy and enstrophy dissipation in 2D
Navier-Stokes flows in the presence of external forcing were obtained
in \cite{AlexakisDoering2006}.

In the present study we construct families of 2D Navier-Stokes flows
which at fixed values of the viscosity $\nu$ locally maximize the
enstrophy dissipation $\chin$ over the prescribed time window $[0,T]$.
These flows are found using methods of numerical optimization to solve
PDE-constrained optimization problems in which the enstrophy
dissipation \eqref{eq:chi} is maximized with respect to the initial
condition $\varphi$ in \eqref{eq:2DNS} subject to certain constraints.
This is a nonconvex optimization problem and we demonstrate that for
every pair $\nu$ and $T$ it admits several branches of locally
maximizing solutions, each corresponding to a distinct dynamic
mechanism for amplification of palinstrophy (which, as is evident from
\eqref{eq:dEdt}, drives the dissipation of enstrophy). Finally, by
assessing the dependence of the maximum enstrophy dissipation
determined in this way for fixed $T$ on the viscosity for decreasing
values of $\nu$, we arrive at interesting new insights about the
sharpness of the different a priori estimates discussed above.

The structure of the paper is as follows: in the next section we
introduce the optimization problem formulated to maximize the
enstrophy dissipation whereas in Section \ref{sec:optimization} we
outline our gradient-based approach to finding families of local
maximizers of that problem; computational results are presented in
Section \ref{sec:results} whereas discussion and final conclusions are
deferred to the last section.

\section{Optimization Problem} 
\label{sec:problem}

Given a fixed viscosity $\nu$ and length $T$ of the time window, we
aim to construct flows maximizing the enstrophy dissipation $\chin$
which will be accomplished by finding suitable optimal initial
conditions ${\phichk}$ in system \eqref{eq:2DNS}.  Since the enstrophy
dissipation is given in terms of a time integral of the palinstrophy,
cf.~\eqref{eq:chi}, we will restrict our attention to initial data
with bounded palinstrophy $\P_0 := \P(\varphi)$, even though system
\eqref{eq:2DNS} admits classical solutions for a much broader class of
initial data \cite{kl04}. We thus have the following optimization
problem.
\begin{problem}
\label{pb:maxchi}  
Given {$\P_0,\nu, T > 0$} in system \cref{eq:2DNS} and the
objective functional \eqref{eq:chi}, find
\begin{equation*}
{\phichk} = \underset{\varphi \in \mathcal{S}} {\argmax} \, \chin(\varphi), \quad \textrm{where} 
\quad \mathcal{S} := \left\{ \varphi \in H^1(\Omega) \: : \: \intO \varphi(\x) \, d\x = 0, \  \P(\varphi) = \P_0 \right\}. 
\end{equation*}
\end{problem}
The Sobolev space $H^1(\Omega)$ is endowed with the inner product
\begin{equation}
\forall_{p_1,p_2 \in H^1(\Omega)} \qquad
\left\langle p_1, p_2 \right\rangle_{H^1(\Omega)} 
 = \intO p_1  p_2 + \ell^2 \, \bfgrad p_1 \cdot \bfgrad p_2 \, d\x, 
\label{eq:ipH1}
\end{equation}
where $\ell \in \RR^+$ is a parameter. We note that the inner products
in \eqref{eq:ipH1} corresponding to different values of $\ell$ are
equivalent as long as $0 < \ell < \infty$. However, as will be shown
in the next section, the choice of the parameter $\ell$ plays an
important role in the numerical solution of Problem \ref{pb:maxchi}. With the initial palinstrophy $\P_0$ fixed, we will find families of
locally maximizing solutions of Problem \ref{pb:maxchi} parameterized
by $T$ for a range of viscosities $\nu$. Our approach to finding such
local maximizers is described next.


\section{Solution Approach} 
\label{sec:optimization}

\subsection{Gradient-Based Optimization}

Since Problem \ref{pb:maxchi} is designed to test certain subtle
mathematical properties of system \eqref{eq:2DNS}, we choose to
formulate the solution approach in the continuous
(``optimize-then-discretize'') setting, where the optimality
conditions, constraints and gradient expressions are derived based on
the original PDE before being discretized for the purpose of numerical
evaluation, instead of the alternative ``discretize-then-optimize''
approach often used in applications \cite{g03}. We first describe the
discrete gradient flow focusing on computation of the gradient of the
objective functional $\chin(\varphi)$ with respect to the initial
condition $\varphi$ and then provide some details about numerical
approximations.

For given values of $\P_0$, $\nu$ and $T$, a local maximizer ${\phichk}$
of Problem \ref{pb:maxchi} can be found as ${\phichk} =
\lim_{n\rightarrow \infty} \varphi^{(n)}$ using the following
iterative procedure representing a discretization of a gradient flow
projected on $\mathcal{S}$
\begin{equation}
\begin{aligned}
\varphi^{(n+1)} & =  \PP_{\mathcal{S}}\left(\;\varphi^{(n)} + \tau_n \nabla\chin\left(\varphi^{(n)}\right)\;\right), \\ 
\varphi^{(1)} & =  \varphi_0,
\end{aligned}
\label{eq:desc}
\end{equation}
where $\varphi^{(n)}$ is an approximation of the maximizer obtained at
the $n$-th iteration, $\varphi_0$ is the initial guess assumed to have
zero mean and $\tau_n$ is the length of the step in the direction of
the gradient $\nabla\chin(\varphi^{(n)})$.  The palinstrophy
constraint is enforced by application of a projection operator
$\PP_{\mathcal{S}} \; : \; H^1(\Omega) \rightarrow
\mathcal{S}$ to be defined below.

A key step in procedure \eqref{eq:desc} is evaluation of the gradient
$\nabla\chin(\varphi)$ of the objective functional $\chin(\varphi)$,
cf.~\eqref{eq:chi}, representing its (infinite-dimensional)
sensitivity to perturbations of the initial condition $\varphi$, and
it is essential that the gradient be characterized by the required
regularity, namely, $\nabla\chin(\varphi) \in H^1(\Omega)$.  This is,
in fact, guaranteed by the Riesz representation theorem \cite{l69}
applicable because the G\^ateaux (directional) differential
$\chin'(\varphi;\cdot) : H^1(\Omega) \rightarrow \RR$, defined as
$\chin'(\varphi;\varphi') := \lim_{\epsilon \rightarrow 0}
\epsilon^{-1}\left[\chin(\varphi+\epsilon \varphi') -
  \chin(\varphi)\right]$ for some perturbation $\varphi' \in
H^1(\Omega)$, is a bounded linear functional on $H^1(\Omega)$.  The
G\^ateaux differential can be computed directly to give
\begin{align}
\chin'(\varphi; \varphi') =& {\frac{2\nu}{T}} \, \int_{0}^{T} \intO  \bfgrad\wn(\x, t; \varphi) \cdot  \bfgrad\wn'(\x, t; \varphi, \varphi')\, d\x dt \nonumber \\
=& -{\frac{2\nu}{T}} \, \int_{0}^{T} \intO  \bfDelta\wn(\x, t; \varphi) \, \wn'(\x, t; \varphi, \varphi') \, d\x dt, 
\label{eq:dchin}
\end{align}
where the last equality follows from integration by parts and  the
perturbation field $\wn' = \wn'(\x, t; \varphi, \varphi')$ is a
solution of the Navier-Stokes \eqref{eq:2DNS} system linearized around
the trajectory corresponding to the initial data $\varphi$ \cite{g03},
i.e.,
\begin{subequations}
\label{eq:2DPert}
\begin{align}
 \K\begin{bmatrix} \wn' \\ \psin' \end{bmatrix} & := 
 \begin{bmatrix}
\p{\wn'}{t} + \bfgrad^{\perp} \psin' \cdot \bfgrad \wn + \bfgrad^{\perp} \psin \cdot \bfgrad \wn' - \nu \bfDelta \wn' \\
\bfDelta \psin' + \wn' 
\end{bmatrix} = \begin{bmatrix}  0 \\ 0\end{bmatrix}, \label{eq:Pert_eqn} \\
\wn'(t=0) &= \varphi', \label{eq:Pert_IC}
\end{align}
\end{subequations}
which is subject to the periodic boundary conditions and where
$\psin'$ is the perturbation of the stream function $\psin$. The Riesz
representation theorem then allows us to write
\begin{equation}
\chin'(\varphi;\varphi') 
= \Big\langle \nabla\chin(\varphi), \varphi' \Big\rangle_{H^1(\Omega)}
= \Big\langle \nabla^{L^2}\chin(\varphi), \varphi' \Big\rangle_{L^2(\Omega)},
\label{eq:riesz}
\end{equation}
where the $L^2$ inner product is obtained by setting $\ell = 0$ in
\eqref{eq:ipH1} and the Riesz representers $\nabla\chin(\varphi)$ and
$\nabla^{L^2}\chin(\varphi)$ are the gradients of the objective
functional computed with respect to the $H^1$ and $L^2$ topology,
respectively. We remark that, while the $H^1$ gradient is used
exclusively in the actual computations, cf.~\eqref{eq:desc}, the $L^2$
gradient is computed first as an intermediate step.

However, we note that expression \eqref{eq:dchin} for the G\^{a}teaux
differential is not yet consistent with the Riesz form
\eqref{eq:riesz}, because the perturbation $\varphi'$ of the initial
data \eqref{eq:IC} does not appear in it explicitly as a factor, but
is instead hidden as the initial {condition} in the linearized
problem, cf.~\eqref{eq:Pert_IC}. In order to transform
\eqref{eq:dchin} to the Riesz form, we introduce the {\em adjoint
  states} $\wn^*, \psin^* \; : \; \Omega \times [0,T] \rightarrow \RR$
and the following duality-pairing relation
\begin{equation}
\begin{aligned}
\left( \K\begin{bmatrix} \wn' \\ \psin' \end{bmatrix}, \begin{bmatrix} \wn^* \\ \psin^* \end{bmatrix} \right)
:= & \int_0^T \intO \K\begin{bmatrix} \wn' \\ \psin' \end{bmatrix} \cdot \begin{bmatrix} \wn^* \\ \psin^* \end{bmatrix} \, d\x \, dt  = 0. 
\end{aligned}
\label{eq:dual}
\end{equation}
Performing integration by parts with respect to both space and time in
\eqref{eq:dual} and judiciously defining the {\em adjoint system} as
(also subject to the period boundary conditions)
\begin{subequations}
\label{eq:Adj}
\begin{align}
\K^*\begin{bmatrix} \wn^* \\ \psin^* \end{bmatrix} & := 
 \begin{bmatrix}
-\p{\wn^*}{t} - \bfgrad^{\perp} \psin \cdot \bfgrad \wn^* + \psin^* - \nu \bfDelta \wn^*\\
\bfDelta \psin^* - \bfgrad^{\perp} \cdot (\wn^* \, \bfgrad \wn)
\end{bmatrix} = \begin{bmatrix}  -{\frac{2\nu}{T}} \bfDelta \wn \\ 0\end{bmatrix}, \label{eq:Adj_eqn} \\
\wn^*(t=T) &= 0, \label{eq:Adj_IC}
\end{align}
\end{subequations}
we arrive at 
\begin{equation}
\begin{aligned}
 \left( \K\begin{bmatrix} \wn' \\ \psin' \end{bmatrix}, \begin{bmatrix} \wn^* \\ \psin^* \end{bmatrix} \right) 
& = \left( \begin{bmatrix} \wn' \\ \psin' \end{bmatrix}, \K^*\begin{bmatrix} \wn^* \\ \psin^* \end{bmatrix}\right) - \intO \varphi'(\x) \, \wn^*(\x, 0) \, d\x  \\
& = \underbrace{-{\frac{2\nu}{T}} \, \int_{0}^{T} \intO \wn' \bfDelta\wn \, d\x dt}_{\chin'({\varphi}; \varphi')} - \intO \varphi'(\x) \, \wn^*(\x, 0) \, d\x = 0,
\end{aligned}
\label{eq:dual_Adj}
\end{equation}
where all boundary terms resulting from integration by parts {with
  respect to the space variable} vanish due to periodicity and one of
the terms resulting from integration by parts with respect to time
vanishes as well due to the terminal condition \cref{eq:Adj_IC}.
Identity \eqref{eq:dual_Adj} then implies $\chin'({\varphi}; \varphi')
= \intO \, \varphi'(\x) \, \wn^*(\x, 0) \, d\x$, from which we deduce
the following expression for the $L^2$ gradient, cf.~\eqref{eq:riesz},
\begin{equation} 
\label{eq:gradL2}
\grad^{L^2}\chin(\x) = \wn^*(\x, 0).
\end{equation}

We note that the $L^2$ gradient does not possess the regularity
required to solve Problem \ref{pb:maxchi}.  Identifying the G\^ateaux
differential \eqref{eq:dchin} with the $H^1$ inner product,
cf.~\eqref{eq:ipH1}, integrating by parts and using \eqref{eq:gradL2},
we obtain the required $H^1$ gradient $\nabla\chi$ as a solution of
the elliptic boundary-value problem
\begin{equation}
\left[ \Id \, - \,\ell^2 \,\Delta \right] \nabla\chin
= \nabla^{L^2} \chin  \qquad \text{in} \ \Omega 
\label{eq:gradH1}
\end{equation}
subject to the periodic boundary conditions.  As shown in
\cite{pbh04}, extraction of gradients in spaces of smoother functions
such as $H^1(\Omega)$ can be interpreted as low-pass filtering of the
$L^2$ gradients with parameter $\ell$ acting as the cut-off
length-scale. The value of $\ell$ can significantly affect the rate of
convergence of the iterative procedure \eqref{eq:desc}.

We define the inverse Laplacian on $\Omega$ such that it returns a
zero-mean function. This ensures that the solution $\wn^*$ of the
adjoint system \eqref{eq:Adj} preserves the zero-mean property which
is then also inherited by the $L^2$ and $H^1$ gradients,
cf.~\eqref{eq:gradL2}--\eqref{eq:gradH1}. The projection operator in
\eqref{eq:desc} is then defined in terms of the normalization
(retraction)
\begin{equation} 
\PP_{\mathcal{S}}(\varphi)  = \sqrt{\frac{\P_0}{\P\left(\varphi\right)}}\,\varphi.
\label{eq:PS}
\end{equation}
An optimal step size $\tau_n$ can be determined by solving the
minimization problem
\begin{equation}
\tau_n = \underset{\tau > 0}{\argmax}\left\{ \chin\left( \mathbb{P}_{\mathcal{S}} \left( \varphi^{(n)} + \tau \, \grad\chin(\varphi^{(n)}) \right) \right) \right\},
\label{eq:tau_n}
\end{equation}
which can be interpreted as a modification of a standard line search
problem with optimization performed following an arc (a geodesic)
lying on the constraint manifold $\mathcal{S}$, rather than a straight
line.

To summarize, a single iteration of the gradient algorithm
\eqref{eq:desc} requires solution of the Navier-Stokes system
\eqref{eq:2DNS} followed by the solution of the adjoint system
\eqref{eq:Adj}, which is a terminal-value problem and hence needs to
be integrated backward in time whereas its coefficients are determined
by the solution of the Navier-Stokes system obtained before. These two
solves allow one to evaluate the $L^2$ gradient via \eqref{eq:gradL2}
which is then ``lifted'' to the space $H^1$ by solving
\eqref{eq:gradH1}. Finally, the approximation of the optimal initial
condition ${\phichk}$ is updated using \eqref{eq:desc} with the step
size $\tau_n$ determined in \eqref{eq:tau_n}. As a first initial guess
$\varphi_0$ in \cref{eq:desc} we use the initial condition constructed
in \cite{Jeong2021} and then, to ensure the maximizers ${\phichk}$
obtained for the same viscosity $\nu$ but different lengths $T$ of the
time window lie on the same {maximizing} branch, we use a continuation
approach where the maximizer ${\phichk}$ is {employed} as the initial
guess $\varphi_0$ to compute
$\widecheck{{\varphi}}_{\nu}^{T+\Delta T}$ for some sufficiently small
$\Delta T$.  We refer the reader to \cite{ap16} for further details of
the continuation approach.

\subsection{Computational Approach}

Both the Navier-Stokes \cref{eq:2DNS} and the corresponding adjoint
system \cref{eq:Adj} are discretized in space using a standard Fourier
pseudo-spectral method. Evaluation of nonlinear products and terms
with nonconstant coefficients is performed using the $2/3$ rule
combined with a Gaussian filter defined by
$\rho(\kk) = e^{-36 \big(\frac{|\kk|}{K} \big)^{36}}$, where $\kk$ is
the wavenumber, $K = \frac{2N}{3}$ and $N$ is the number of Fourier
modes used in each direction \cite{Hou2009}. Time integration is
carried out using a four-step, globally third-order accurate mixed
implicit/explicit Runge-Kutta scheme with low truncation error
\cite{Alimo2020}. The results presented in the next section were
obtained using the spatial resolutions $N = 512, 1024$ and the
time-steps
$\Delta t \approx 4.4721 \times 10^{-5}, 2.2361 \times 10^{-5}, 8.9443
\times 10^{-6}$, with finer resolutions employed for problems with
smaller values of the viscosity $\nu$. In system \eqref{eq:gradH1}
defining the Sobolev gradients we set $\l =1$ and a spectral method is
used to solve this system.  The line-search problem \eqref{eq:tau_n}
is solved with Brent's derivative-free algorithm
\cite{press2007numerical}.  Due to its large computational cost, a
massively parallel implementation of the approach presented has been
developed in FORTRAN 90 using the Message Passing Interface (MPI).

\section{Results}
\label{sec:results}

In this section we present the results obtained by solving Problem
\ref{pb:maxchi} with $\P_0 = 1$ fixed and both $\nu$ and $T$ varying
over a broad range of values.  In addition to understanding the
structure of the flows maximizing the enstrophy dissipation and how it
changes when the parameters are varied, our goal is also to provide
insights which of the estimates \eqref{eq:Tran}--\eqref{eq:chiS} best
describe the behavior of the maximum enstrophy dissipation
$\chin(\phichk)$ in the limit of vanishing viscosity.

Problem \ref{pb:maxchi} is nonconvex and as such admits multiple local
maximizers at least for some values of $\nu$ and $T$. Information
about the six distinct local maximizers found for
$\nu = 2.2361 \times 10^{-6}$ and $T = 0.1789$ is collected in Table
\ref{tab:branches} where we show the corresponding palinstrophy
evolutions $\P(t)$, optimal initial conditions $\phichk(\x)$ and the
vorticity fields realizing the maximum palinstrophy
$\wn(\x,\argmax_{0< t \le T}\P(t))$. The time evolution of the
vorticity fields corresponding to all six branches is visualized in
\href{https://youtu.be/4I_obQAgxUY}{Movie 1}. This movie offers
insights about the different physical mechanisms involving the
stretching of thin vorticity filaments which are responsible for the
growth of palinstrophy and hence also increased enstrophy
dissipation. It is noteworthy that all these flow evolutions feature
very thin filaments which however do not undergo the Kelvin-Helmholtz
instability as they are stabilized by vortices also present in the
flow field. Flows on branches 3 and 4, which feature multiple
palinstrophy maxima, employ a mechanism similar to the continuous
baker's map to amplify the palinstrophy.  Moreover, we see that,
interestingly, in some cases seemingly very similar optimal initial
conditions $\phichk$ give rise to quite different flow evolutions
featuring different numbers of local palinstrophy maxima (one or two)
in the considered time window $[0,T]$, see, e.g., the maximizers from
Branches 2 and 3 in Table \ref{tab:branches}. This makes classifying
local optimizers into branches a rather difficult task and the
classification presented in Table \ref{tab:branches} is tentative
only, which will however not affect the main findings of our
study. \href{https://youtu.be/8b9q360Bumc}{Movie 2} and
\href{https://youtu.be/dMiBYk6YjM0}{Movie 3} show the flow evolutions
and representative palinstrophy histories corresponding to the locally
optimal initial conditions $\phichk$ obtained, respectively, on Branch
1 with $T = 0.1207$ and on Branch 5 with $T = 0.2683$ for five
different values of the viscosity $\nu$. In both cases we see that
even though the optimal initial conditions $\phichk$ obtained for
different values of $\nu$ are quite similar, qualitative changes occur
in the flows evolutions as the viscosity is reduced. We attribute
these changes to either possible bifurcations of the branches
(understood as functions of $\nu$) or to the possibility that the flow
evolutions corresponding to smaller viscosity values belong to some
unclassified branch, underpinning the difficulty mentioned above.
\begin{landscape}
  \thispagestyle{empty}
\begin{table}
	\centering
	\vspace*{-2.5cm} \hspace*{-2.7cm}\begin{tabular}{ |c|c|c|c|c|c|c|} 
		\hline
		\rowcolor{Gray}
		Branch & 1 & 2 & 3 & 4 & 5 & 6  \\ 
		\hline &&&&&& \\ [-1.5em]
		\rotatebox{90}{\hspace{0.7cm}Palinstrophy} &  {\includegraphics[scale=0.25]{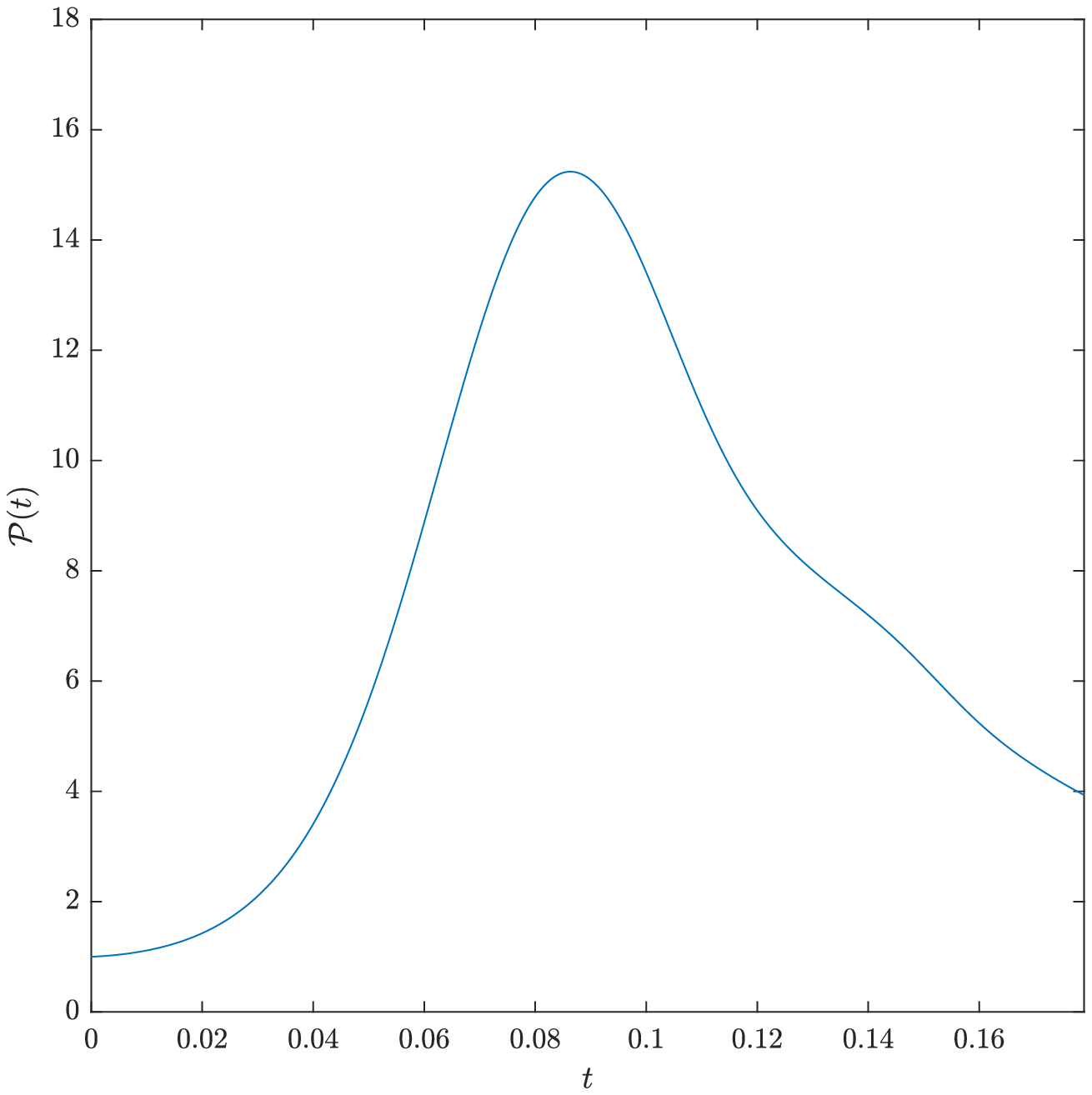}} & 
		{\includegraphics[scale=0.25]{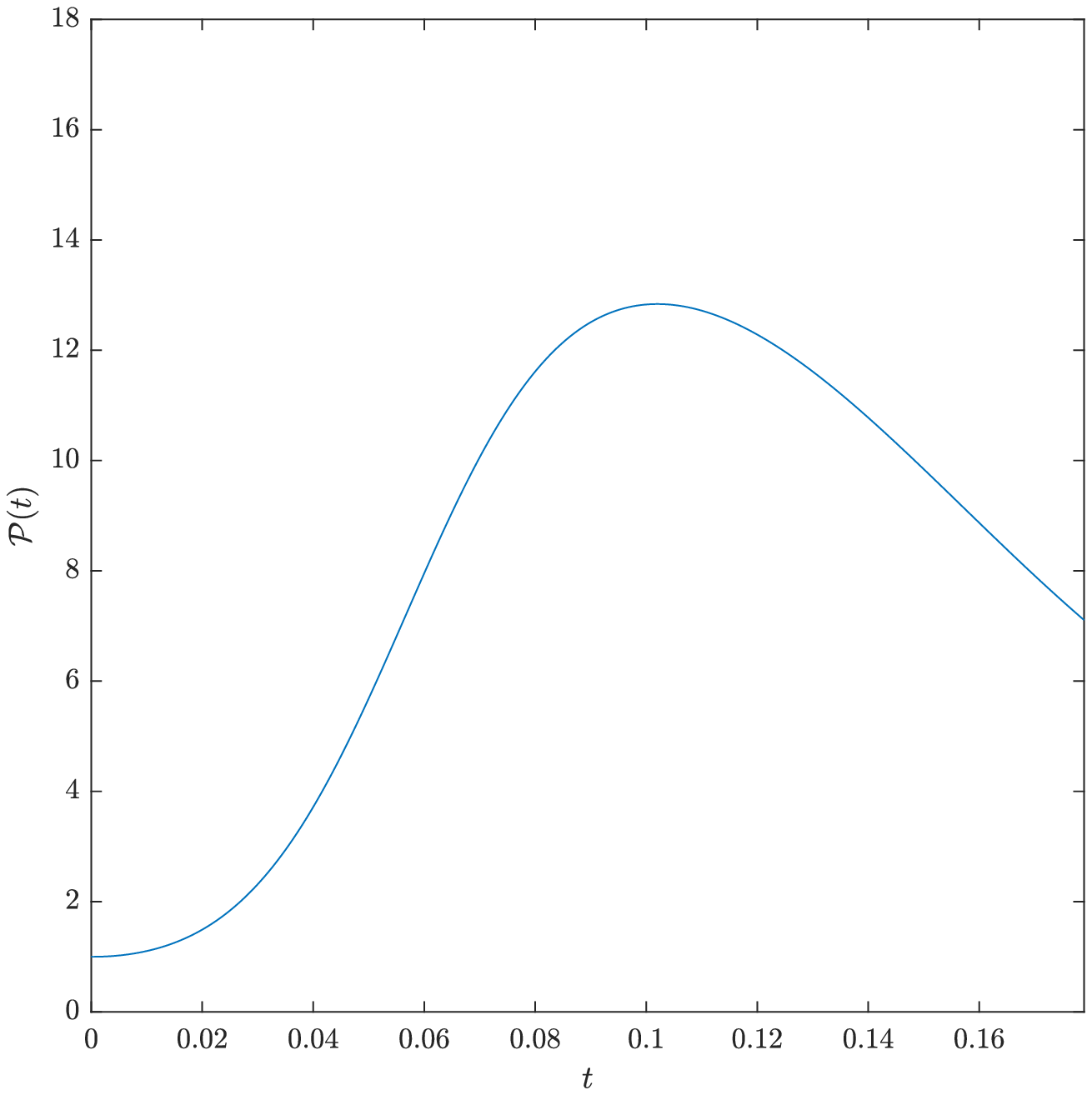}} &
		{\includegraphics[scale=0.25]{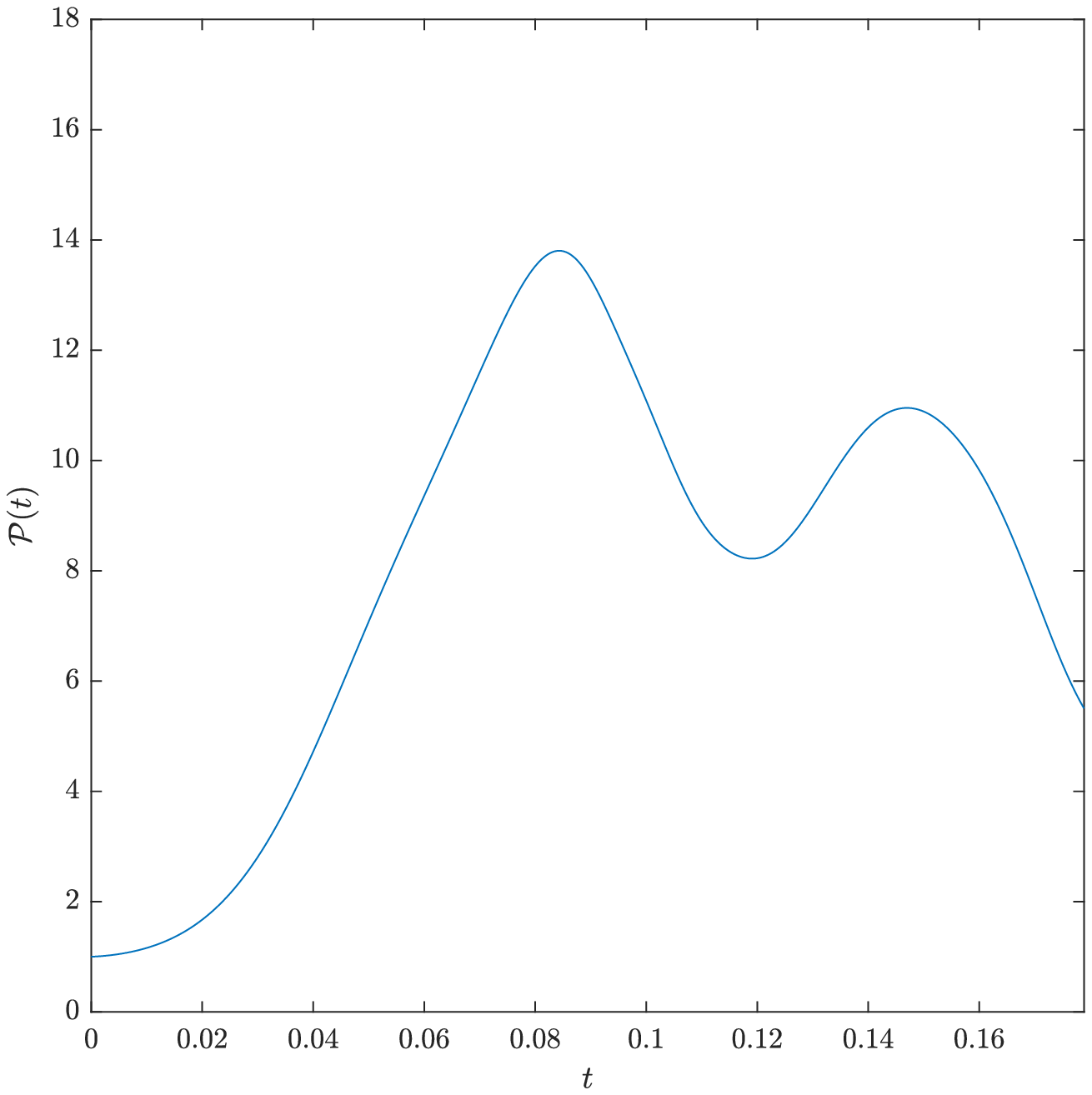}} &
		{\includegraphics[scale=0.25]{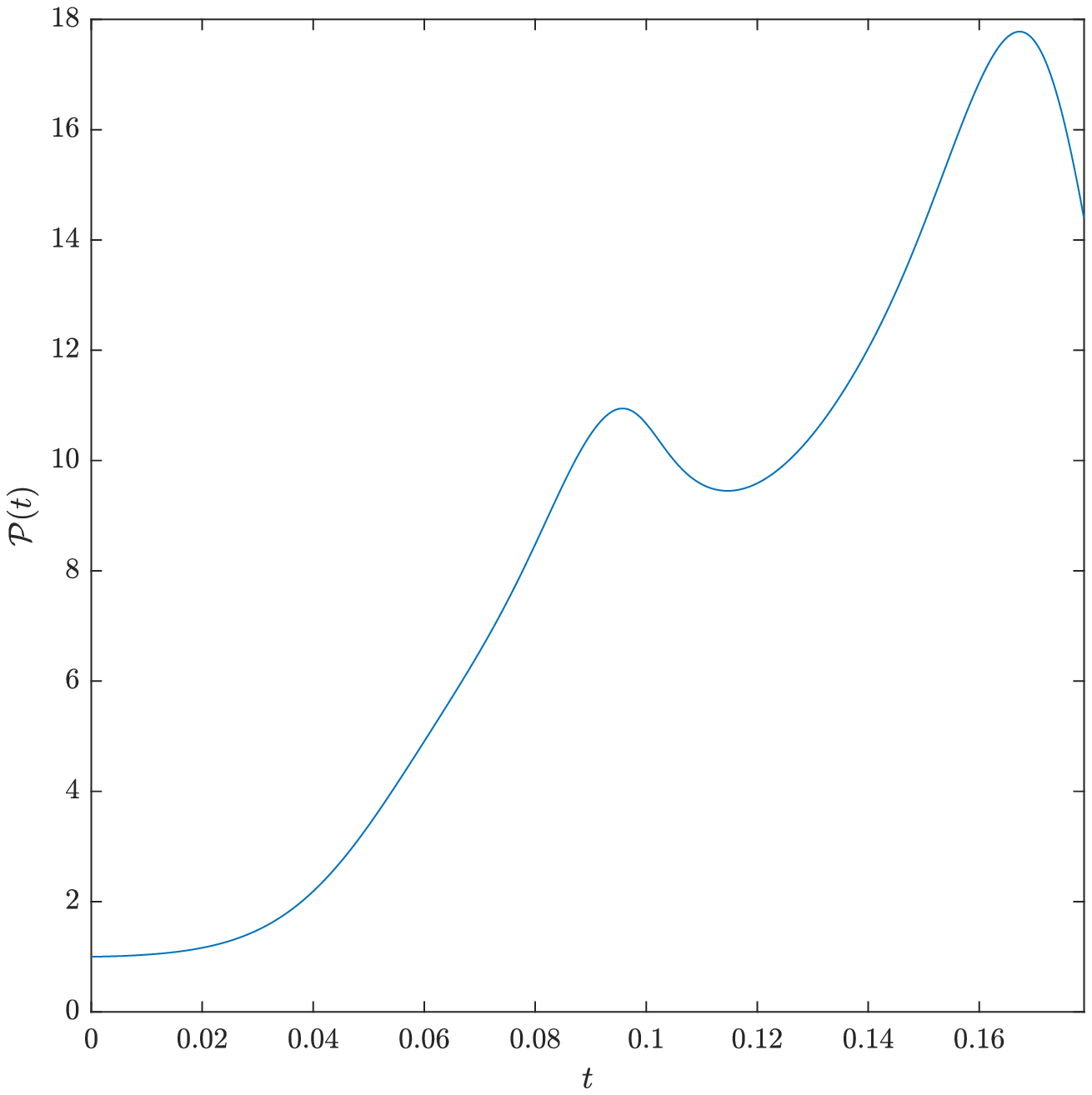}} & 
		{\includegraphics[scale=0.25]{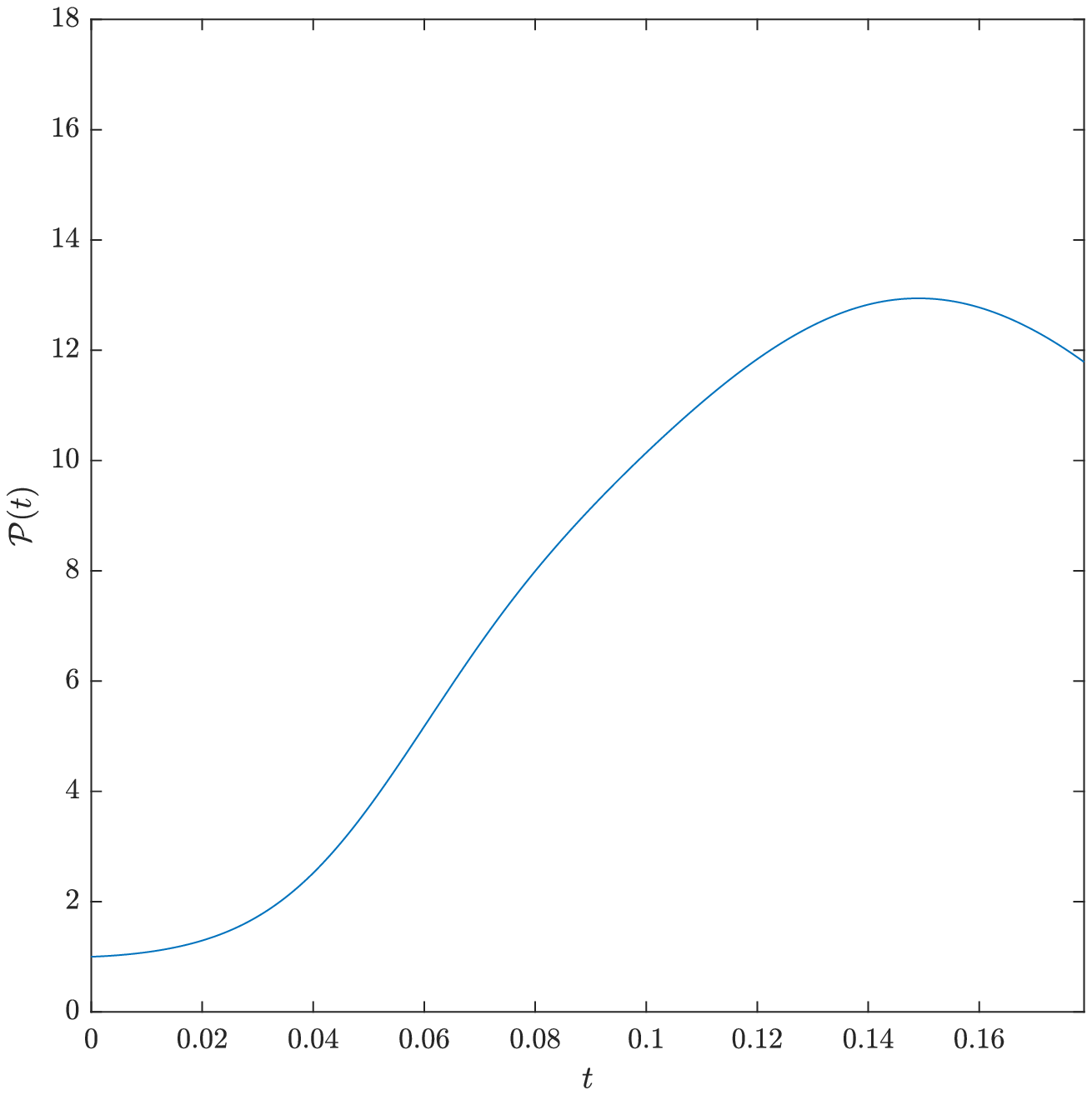}} &
		{\includegraphics[scale=0.25]{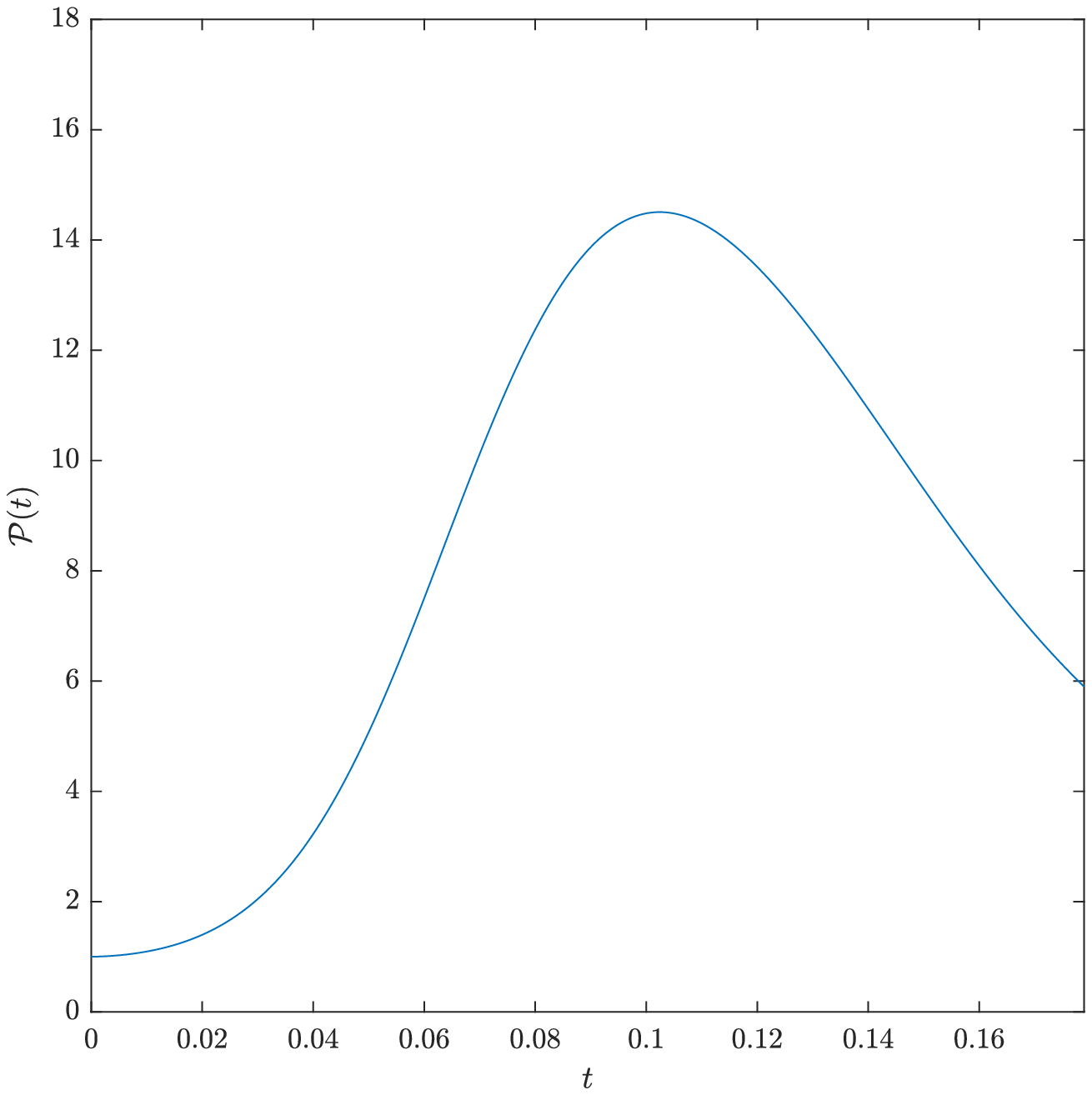}} 
		\\		
		\hline &&&&&& \\ [-1.5em]
		\rotatebox{90}{\hspace{0.6cm}Initial Condition} & {\includegraphics[scale=0.25]{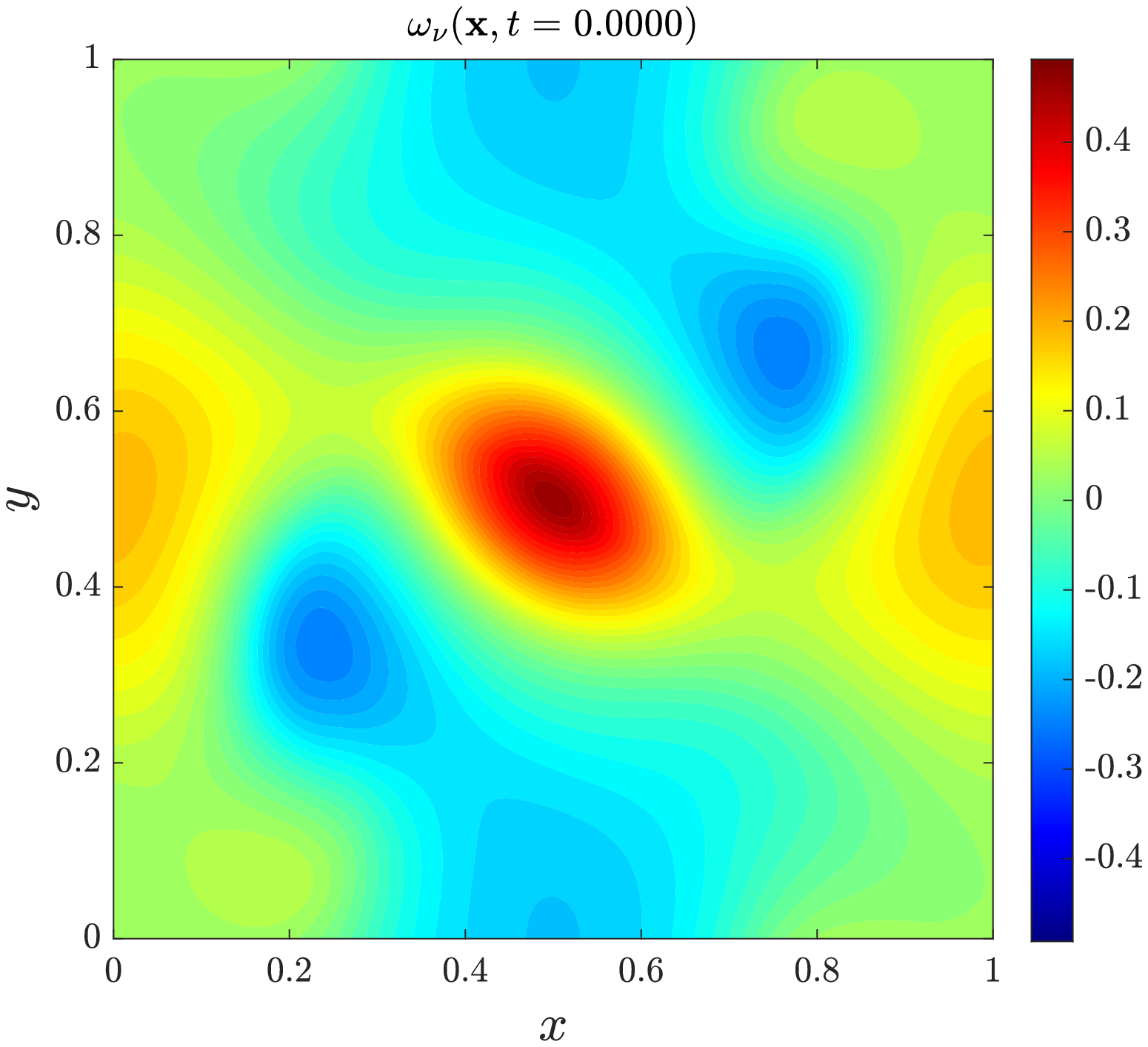}} & 
		{\includegraphics[scale=0.25]{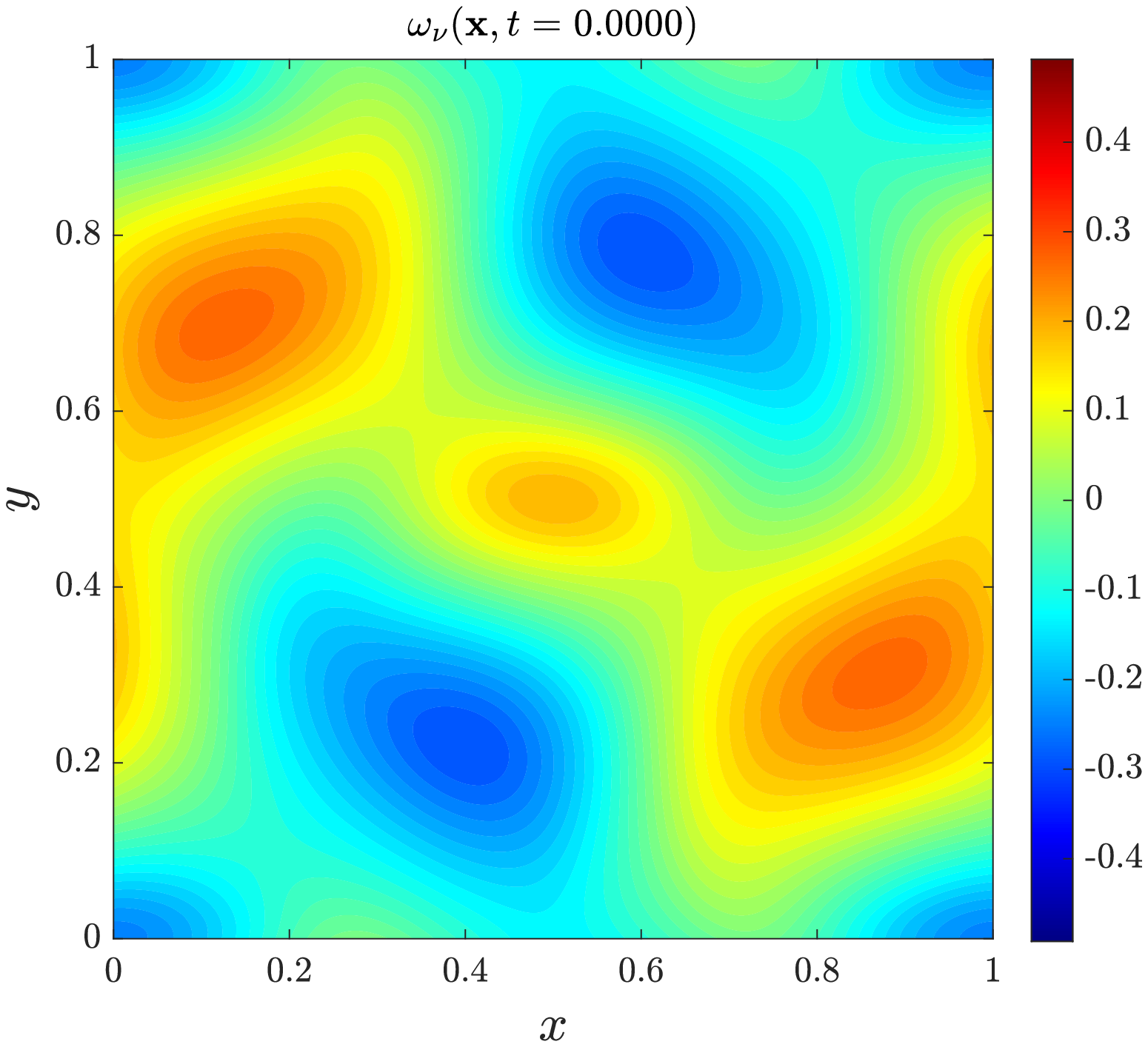}} & 
		{\includegraphics[scale=0.25]{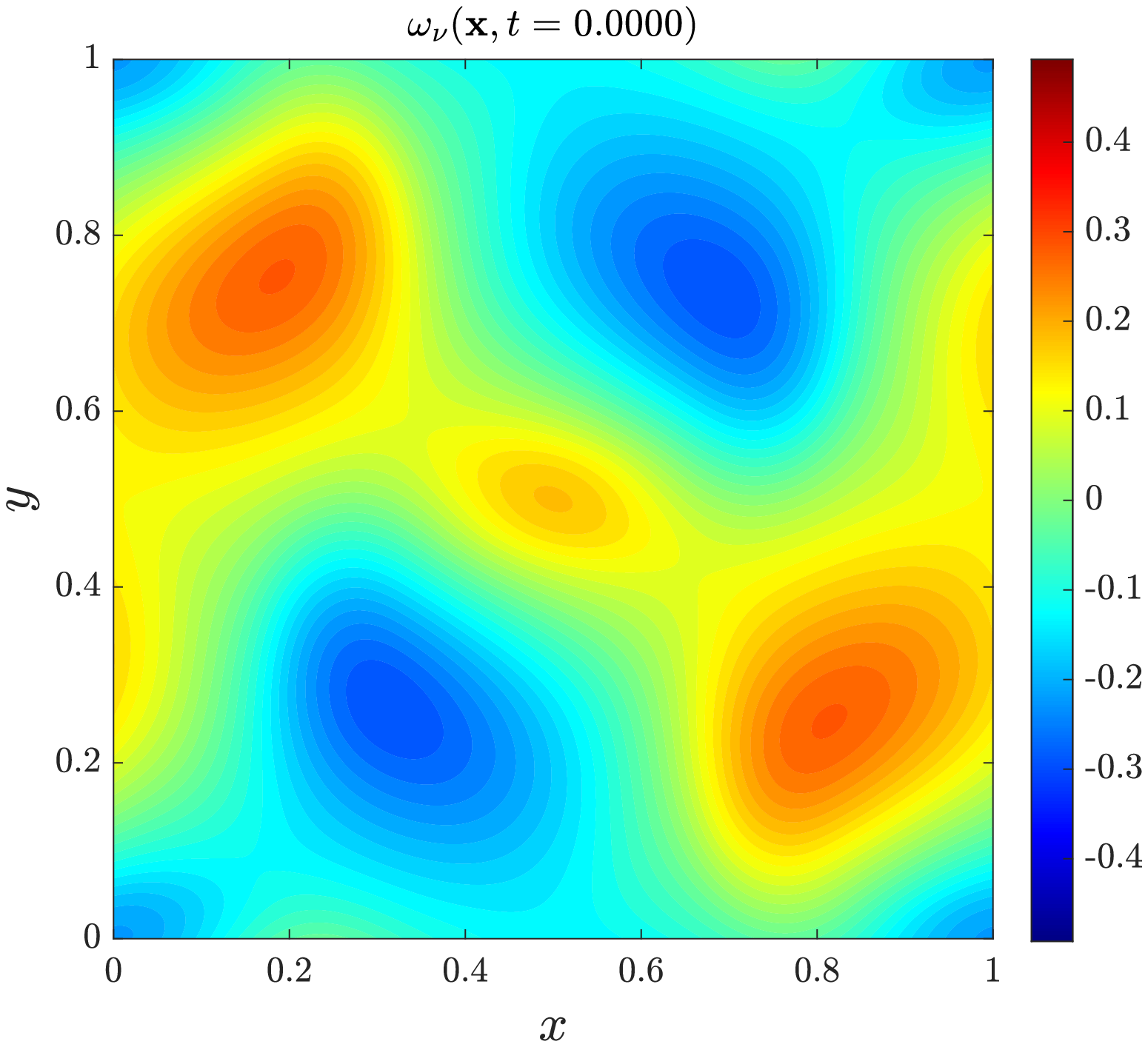}} &
		{\includegraphics[scale=0.25]{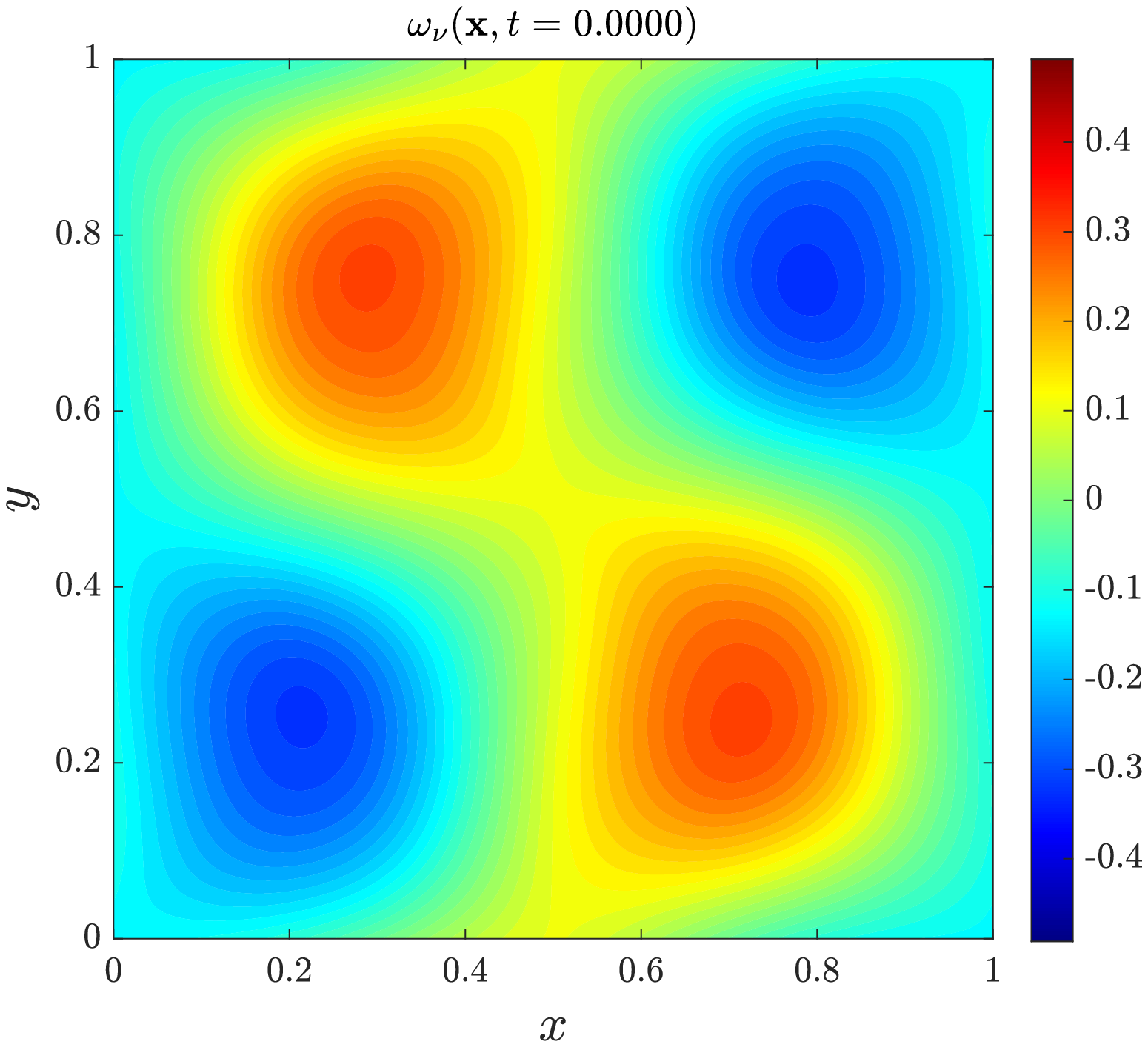}} &
		{\includegraphics[scale=0.25]{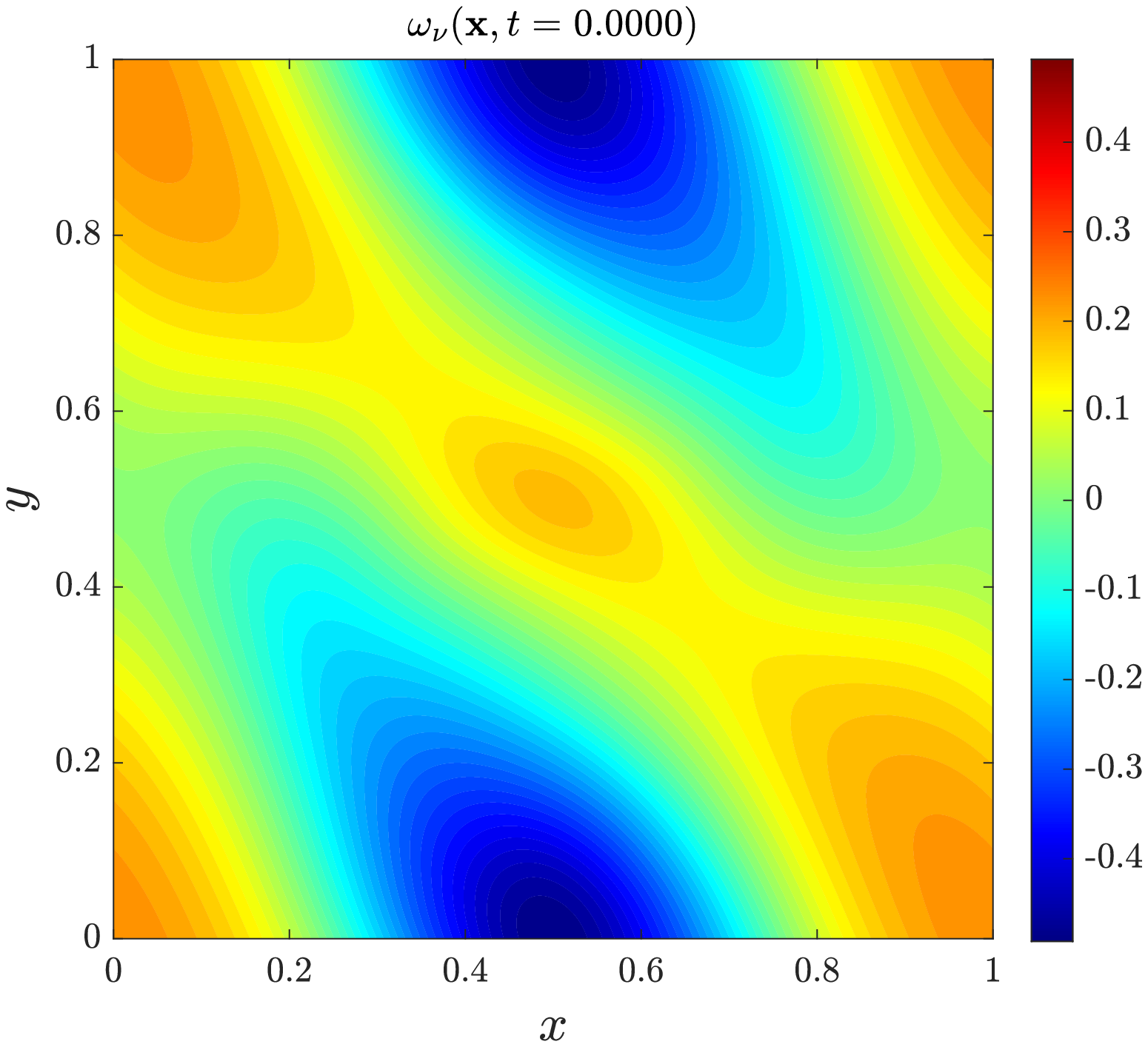}} &
		{\includegraphics[scale=0.25]{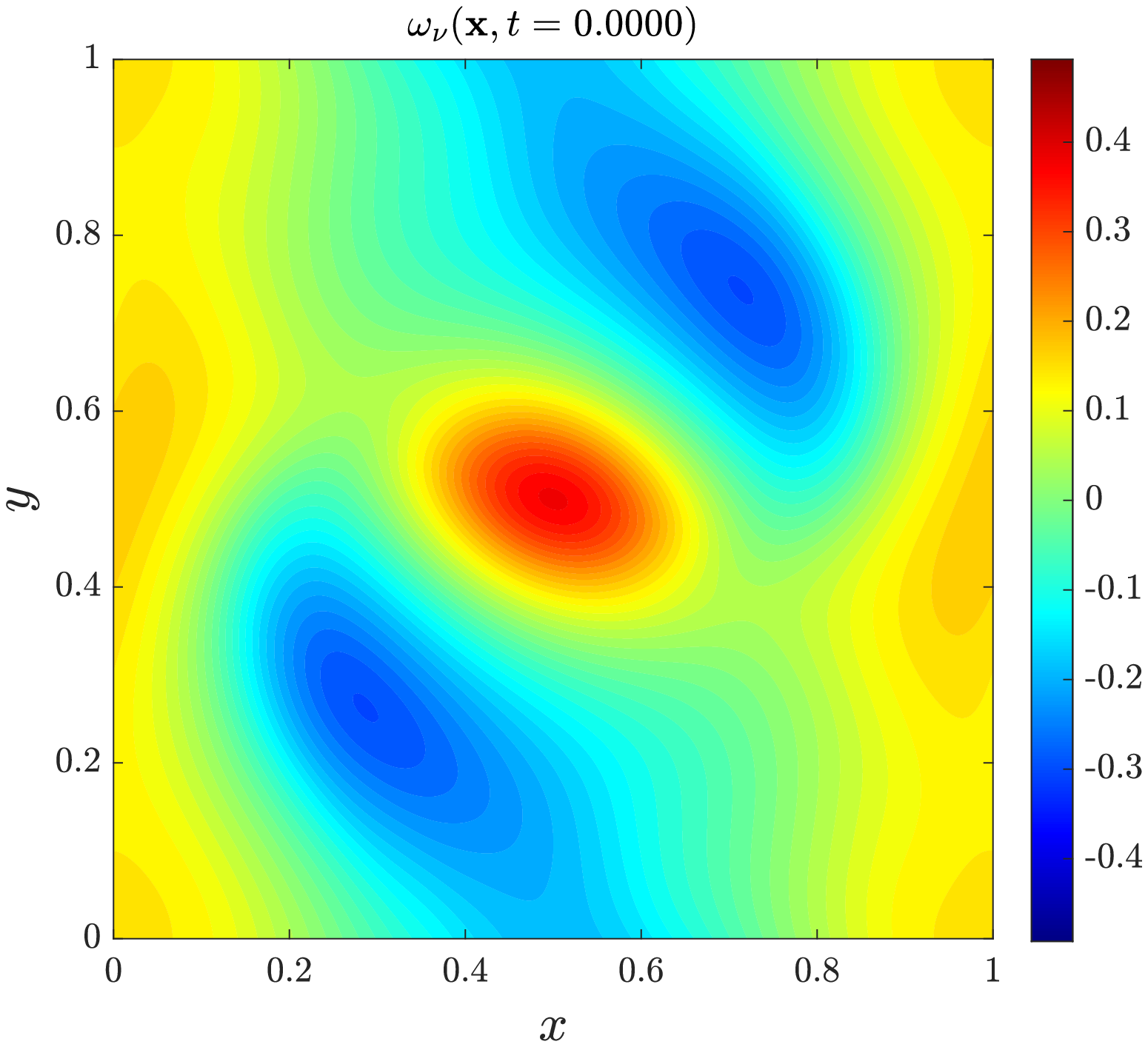}}  \\ 
		\hline &&&&&& \\ [-1.5em]
		\rotatebox{90}{\hspace{0.4cm}Palinstrophy Peak} & 		{\includegraphics[scale=0.25]{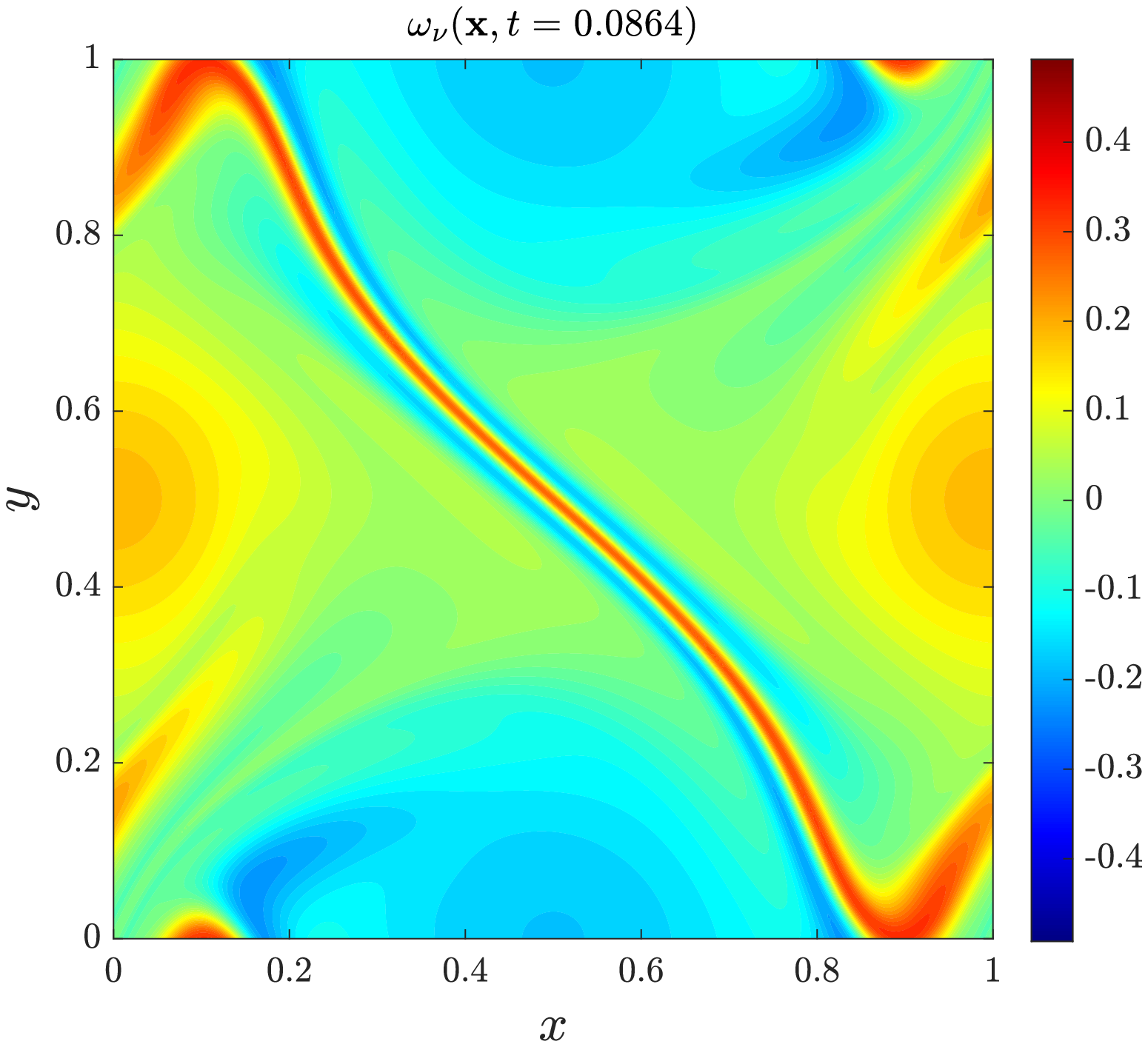}} & 
		{\includegraphics[scale=0.25]{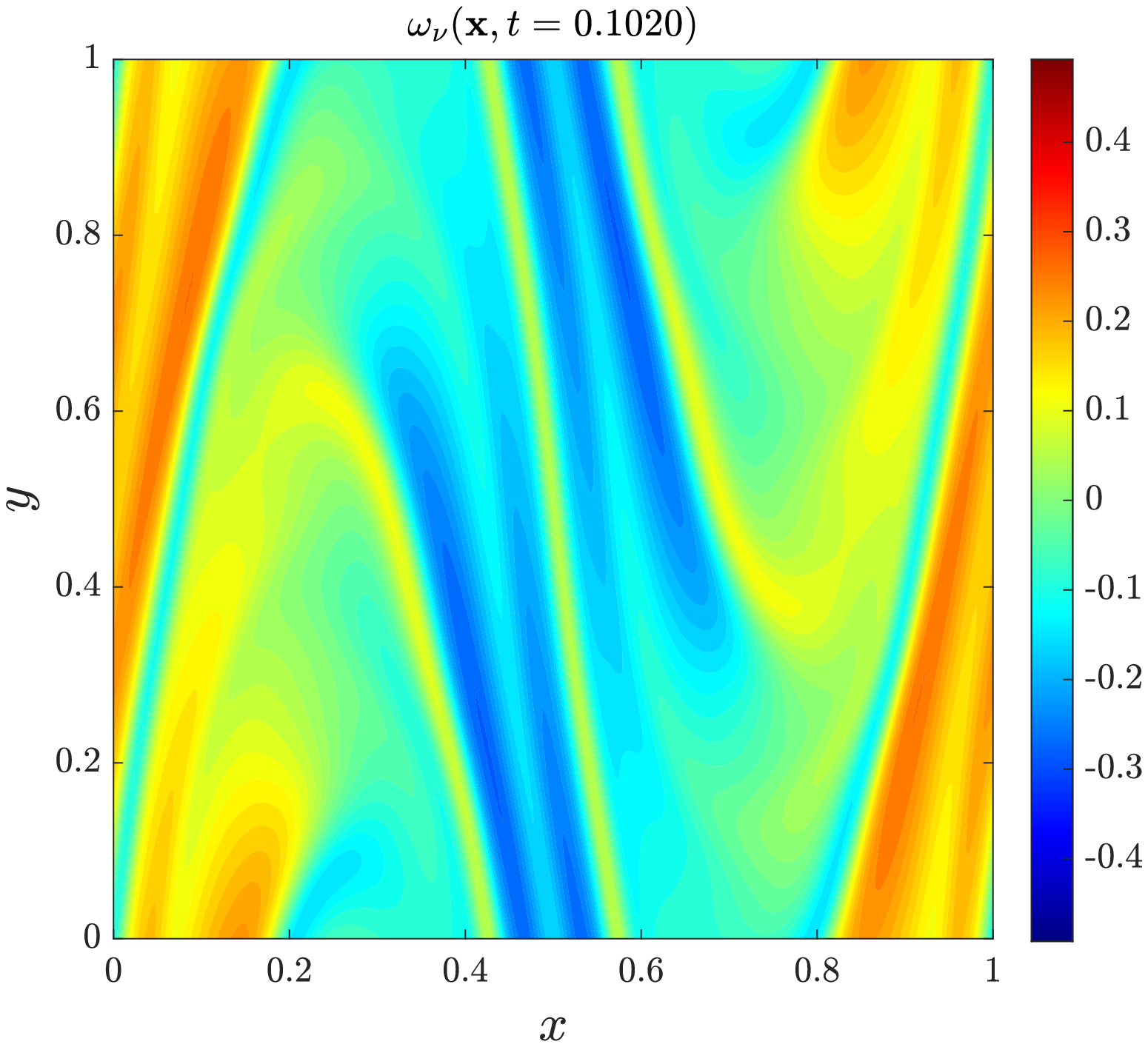}} &
		{\includegraphics[scale=0.25]{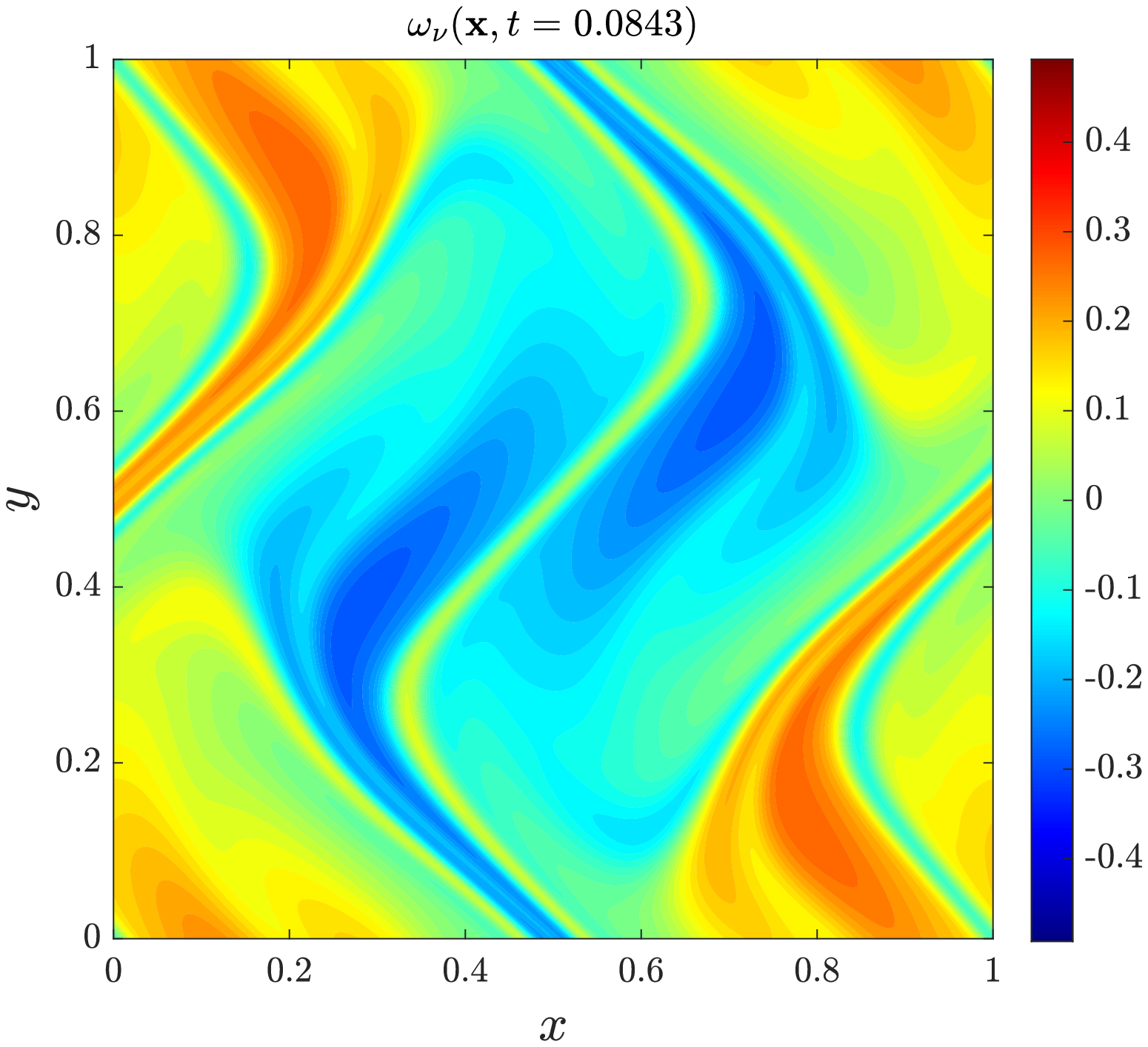}} &
		{\includegraphics[scale=0.25]{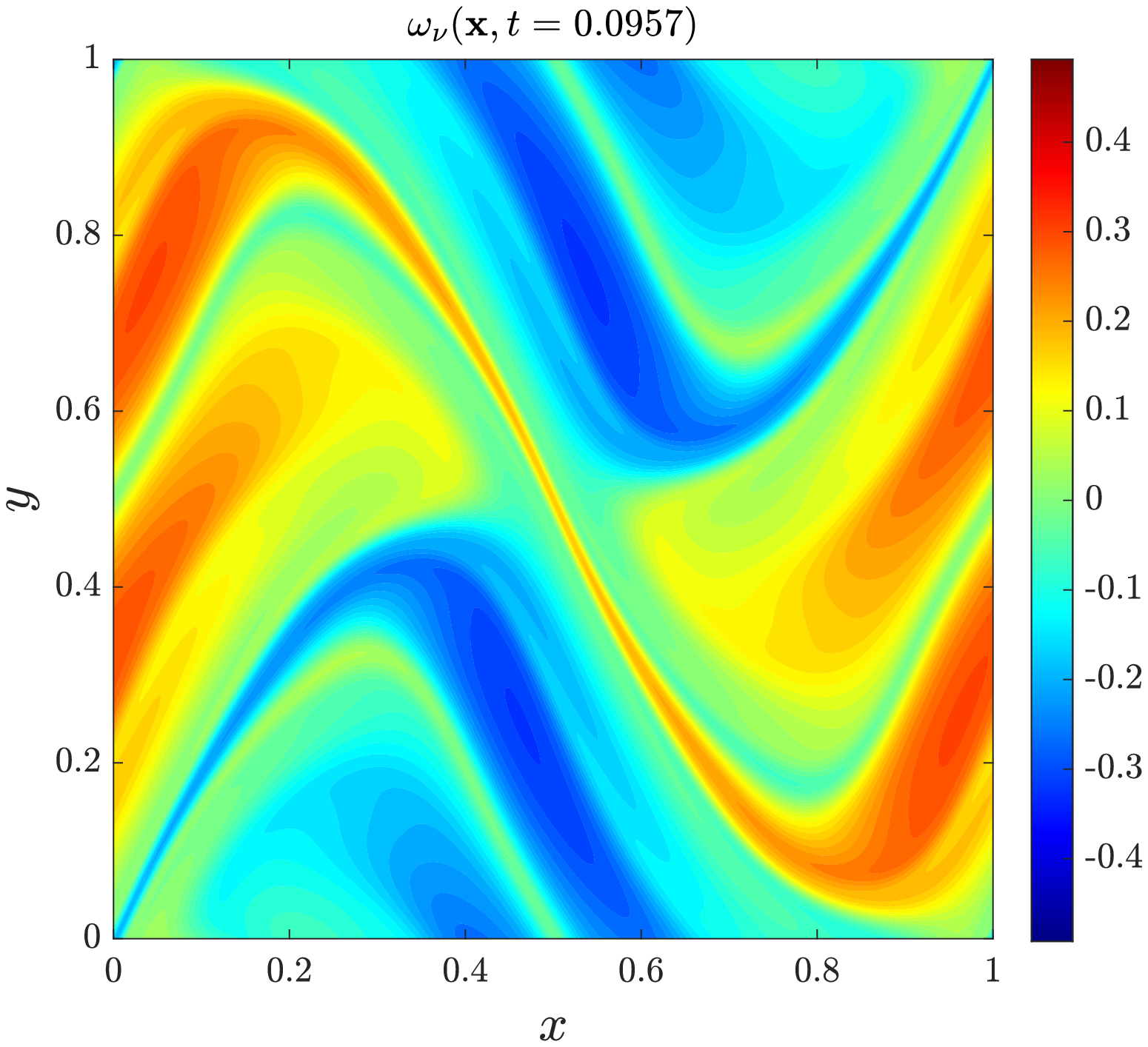}} & 
		{\includegraphics[scale=0.25]{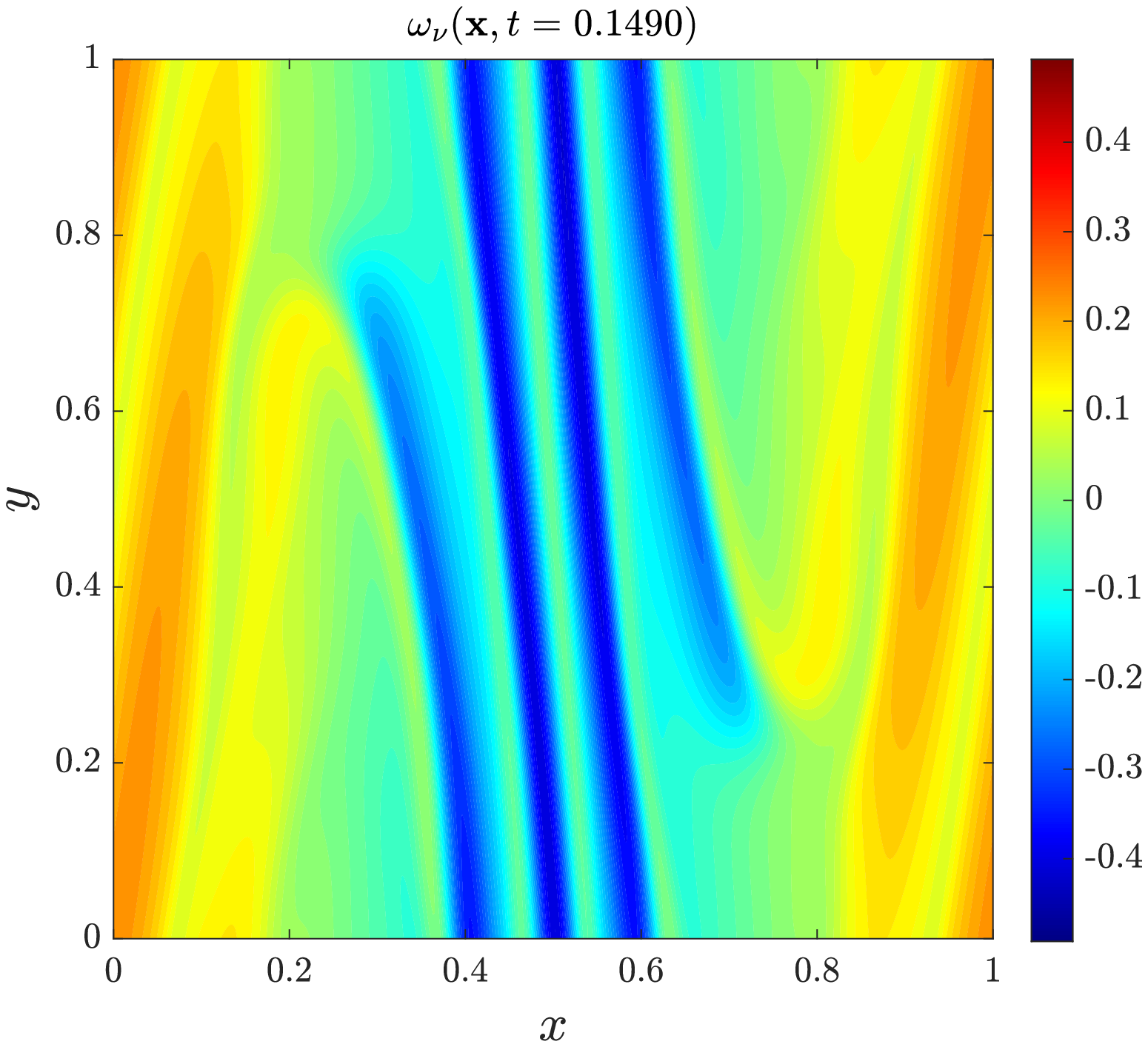}} &
		{\includegraphics[scale=0.25]{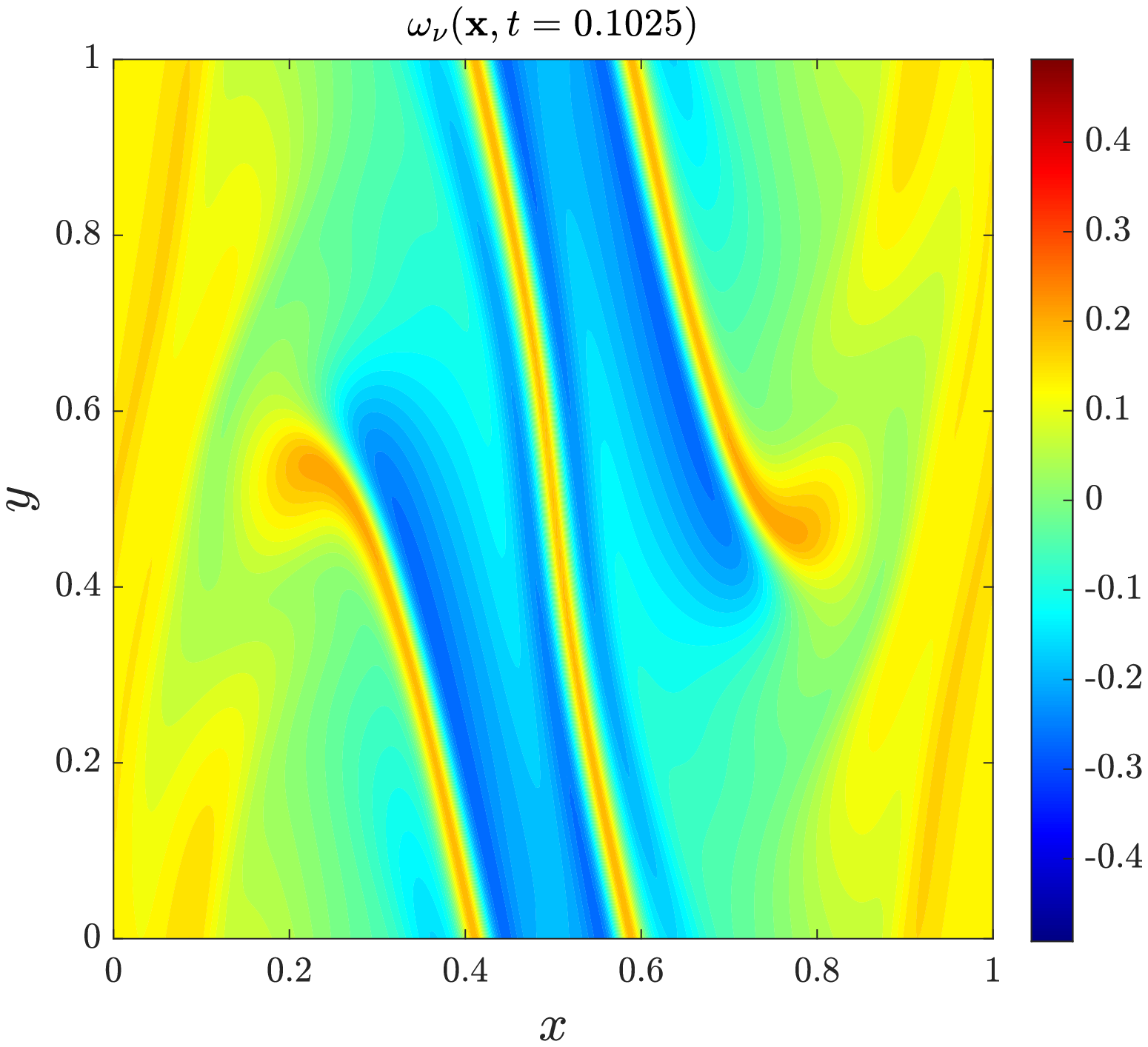}}  \\ 
		\hline &&&&&& \\ [-1.5em]
		\rotatebox{90}{\hspace{0.3cm}Palinstrophy Peak 2} & N/A & N/A &
		\includegraphics[scale=0.25]{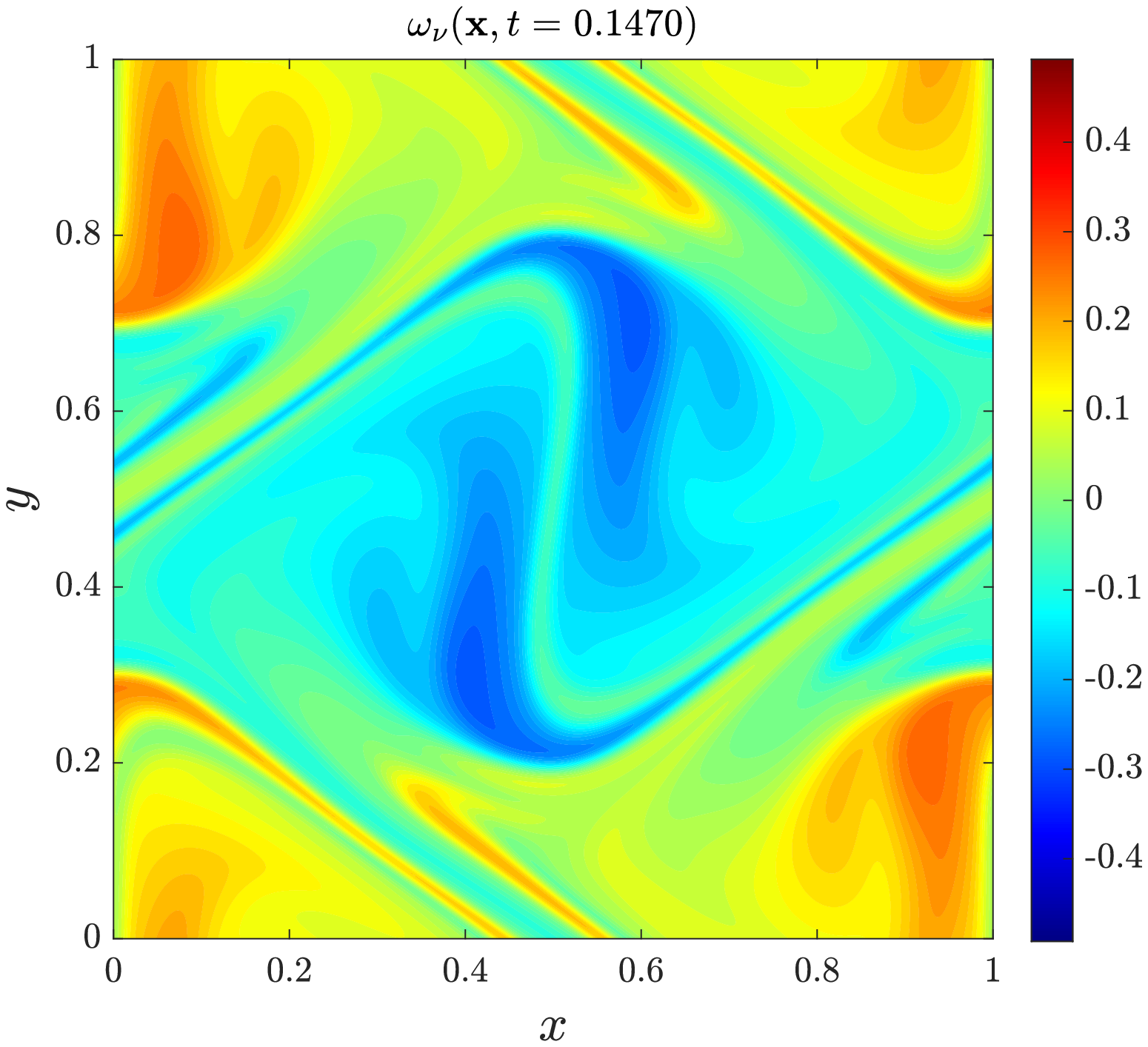} &
		\includegraphics[scale=0.25]{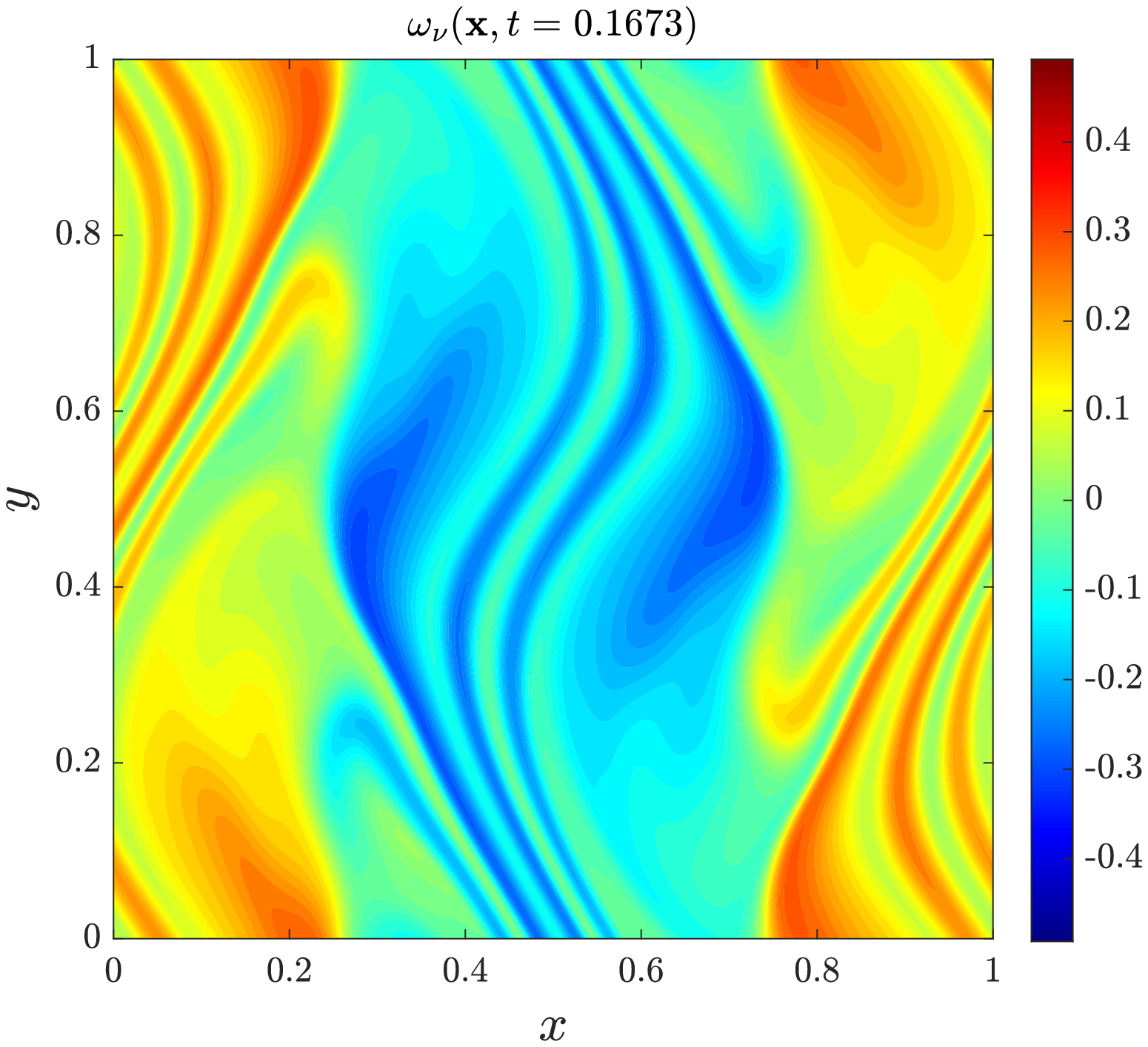} & 
		N/A & 
		N/A 
		\\ 
		\hline
	\end{tabular}
	\caption{Summary information about the local maximizers
          obtained by solving Problem \ref{pb:maxchi} with
          $\nu = 2.2361 \times 10^{-6}$ and $T = 0.1789$. The time
          evolution of the vorticity fields is visualized in
          \href{https://youtu.be/4I_obQAgxUY}{Movie 1}.}
	\label{tab:branches}
\end{table} 
\end{landscape}

\begin{figure}\centering
  \vspace*{-2.0cm}
  \subfigure[]
  {
    \includegraphics[scale=0.725]{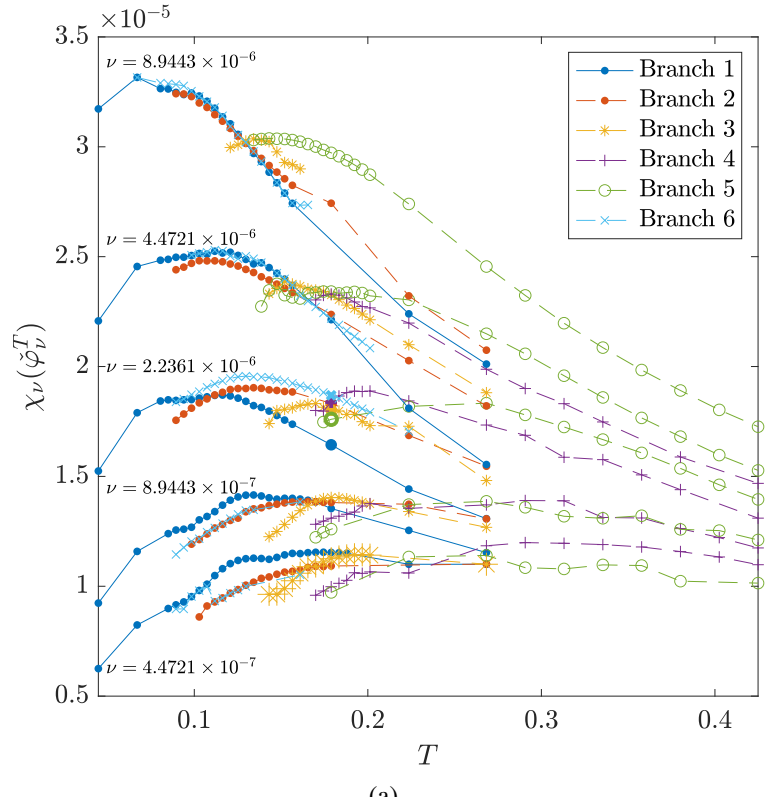}
  }
  \subfigure[]
  {
    \includegraphics[scale=0.725]{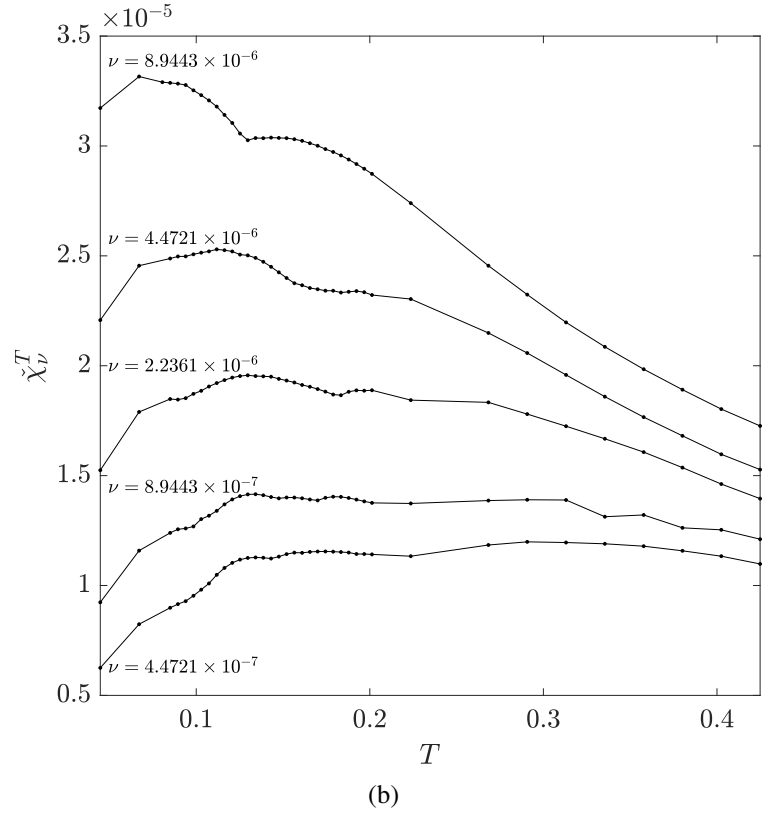}
  }
  \caption{Dependence of (a) the maximum enstrophy dissipation
    $\chin(\phichk)$ for maximizers on the different branches,
    cf.~Table \ref{tab:branches}, and (b) its envelope $\chinck$ on
    the length $T$ of the time window for different viscosities
    $\nu$. In panel (a) the local maximizers illustrated in Table
    \ref{tab:branches} are marked with larger symbols.}
  \label{fig:chiT}
\end{figure}

Next, in Figure \ref{fig:chiT}a we show the dependence of the maximum
enstrophy dissipation $\chin(\phichk)$ on the length $T$ of the time
window for five values of viscosity spanning more than one order of
magnitude. We carefully distinguish branches of distinct local
maximizers, where by a ``branch'' we mean a family of optimal initial
data $\phichk$ parametrized by $T$ and such that the enstrophy
dissipation $\chin(\phichk)$ changes smoothly as $T$ is varied while
$\nu$ remains fixed. We remark that for certain combinations of $\nu$
and $T$ only a subset of the local maximizers described in Table
\ref{tab:branches} could be found. In Figure \ref{fig:chiT}a we
observe that along each branch the maximum enstrophy dissipation
$\chin(\phichk)$ admits a well-defined maximum with respect to $T$. We
add that the values of $\chin(\phichk)$ shown in Figure
\ref{fig:chiT}a are for each value of $\nu$ at least an order of
magnitude larger than the enstrophy dissipation corresponding to the
initial conditions constructed in \cite{Jeong2021}, which realize the
behavior given in \eqref{eq:chiJY}.

As these are the quantities needed to make quantitative comparisons
with estimates \eqref{eq:Tran}--\eqref{eq:chiS}, in Figure
\ref{fig:chiT}b we plot the ``envelopes'', defined as
$\chinck := \max_{\textrm{branches}} \chin(\phichk) $, of the branches
obtained at fixed values of $\nu$. ``Singularities'' evident in these
curves correspond to values of $T$ where different branches become
dominant as $T$ varies.

Next, we move on to identify quantitative connections between the data
presented in Figure \ref{fig:chiT}b and estimates
\eqref{eq:Tran}--\eqref{eq:chiS} describing the vanishing of the
enstrophy dissipation in the inviscid limit $\nu \rightarrow 0$. These
estimates also depend on the length $T$ of the time window, but this
dependence is in some cases less explicit and we will therefore
consider $T$ as a fixed parameter. We thus introduce the following
ans\"atze
\begin{subequations}
  \label{eq:ff}
  \begin{align}
    f_1(\nu) &= C \, \left[-\ln(\nu)\right]^{-\frac{1}{2}}, \label{eq:Tranf} \\
    f_2(\nu) &= C \, \nu^{\alpha}, \label{eq:CCSf} \\
    f_3(\nu) &= C \, \left[\frac{\nu}{|\ln(\nu)|}\right]^{\alpha}, \label{eq:Sf} \\
    f_4(\nu) &= C \, \nu\, \left[-\ln(\nu)\right]^{\frac{1}{2}} \label{eq:JYf}
  \end{align}
\end{subequations}
motivated by the structure of the different estimates. More
specifically, \eqref{eq:Tranf} is the expression from Conjecture
\ref{conj:TD}, cf.~\eqref{eq:Tran}, \eqref{eq:CCSf} has the general
form of the upper bounds in \eqref{eq:chiCDE}--\eqref{eq:chiCCS},
where in the latter case we only consider the second argument of the
function $\max(\cdot)$ since the function $\phi_{\varphi, p, M}$
appearing in the first argument is not given explicitly enough to
allow for quantitative comparisons, \eqref{eq:Sf} is motivated by the
form of estimate \eqref{eq:chiS} whereas \eqref{eq:JYf} is the bound
from Theorem \ref{thm:lower}, cf.~\eqref{eq:chiJY}.

We want to find out which of the functions
\eqref{eq:Tranf}--\eqref{eq:JYf} best describes the dependence of the
data shown in Figure \ref{fig:chiT}b on $\nu$ for different fixed
values of $T$. For each discrete value of $T$ (marked with solid
symbols in Figure \ref{fig:chiT}b) we determine the constant
$C = C(T)$ in each of the ansatz functions
\eqref{eq:Tranf}--\eqref{eq:JYf} by solving the problem
\begin{equation}
{\tC(T) = \argmin_{C \in  \RR^+} \mu_i^T(C),} \qquad  i=1,2,3,4,
\label{eq:C}
\end{equation}
where the least-square error is defined as 
\begin{equation} \label{eq:MSE}
\mu_i^T(C) := \frac{1}{5} \sum_{j=1}^{5} \left[\widecheck{\chi}^T_{\nu_j} - f_i(\nu_j)\right]^2
\end{equation}
with
$\nu_j \in \{ 8.9443 \times 10^{-6}, 4.4721 \times 10^{-6}, 2.2361
\times 10^{-6}, 8.9443 \times 10^{-7}, 4.4721 \times 10^{-7}\}$
representing the considered values of the viscosity.  In addition, we
note that ansatz functions \eqref{eq:CCSf}--\eqref{eq:Sf} also involve
a priori undefined exponents $\alpha \in (0,1)$ and to account for
this fact in each of these cases solution of problem \eqref{eq:C} is
embedded in bracketing procedure which allows us to determine the
exponent $\talpha = \talpha(T)$ producing the smallest error
\eqref{eq:MSE} for a given value of $T$. {The bracketing
  procedure is performed by first determining $\mu_i^T(\tC(T))$ for a
  range of discrete values of $\alpha \in [0,1]$ and then using
  bisection to iteratively improve the approximation of
  $\talpha = \talpha(T)$ which produces the smallest error
  \eqref{eq:MSE}.}  We emphasize that even though the ans\"atze
\eqref{eq:Tranf}--\eqref{eq:JYf} involve different numbers of
parameters (one or two), they are all fitted to the data in Figure
\ref{fig:chiT}b in the same way (i.e., by adjusting $C = C(T)$), which
is done independently for different discrete exponents $\alpha$ in the
case of relations \eqref{eq:CCSf}--\eqref{eq:Sf}.

\begin{figure}\centering
\mbox{\subfigure[]
  {\includegraphics[width=0.5\textwidth]{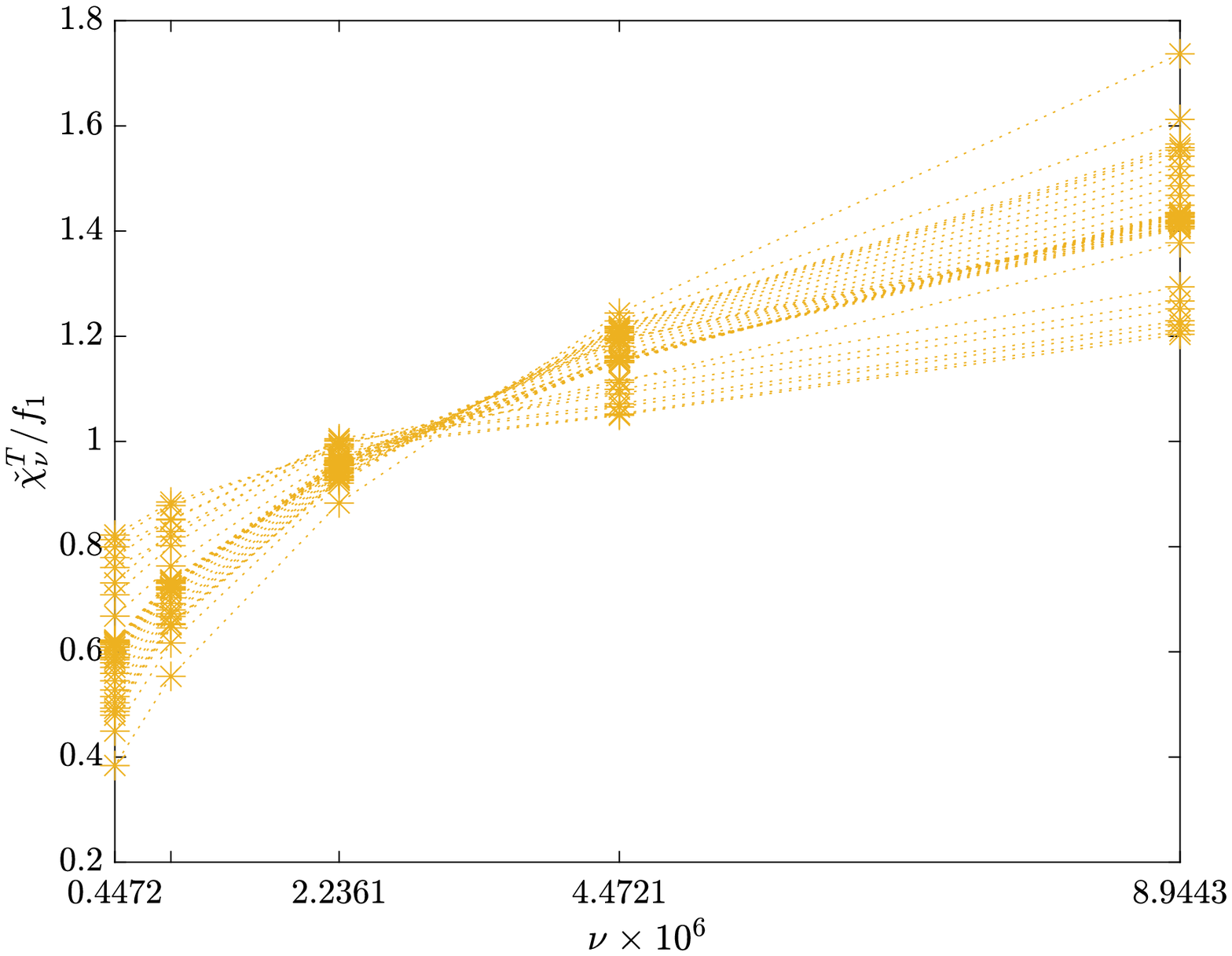}}\qquad
  \subfigure[]
  {\includegraphics[width=0.5\textwidth]{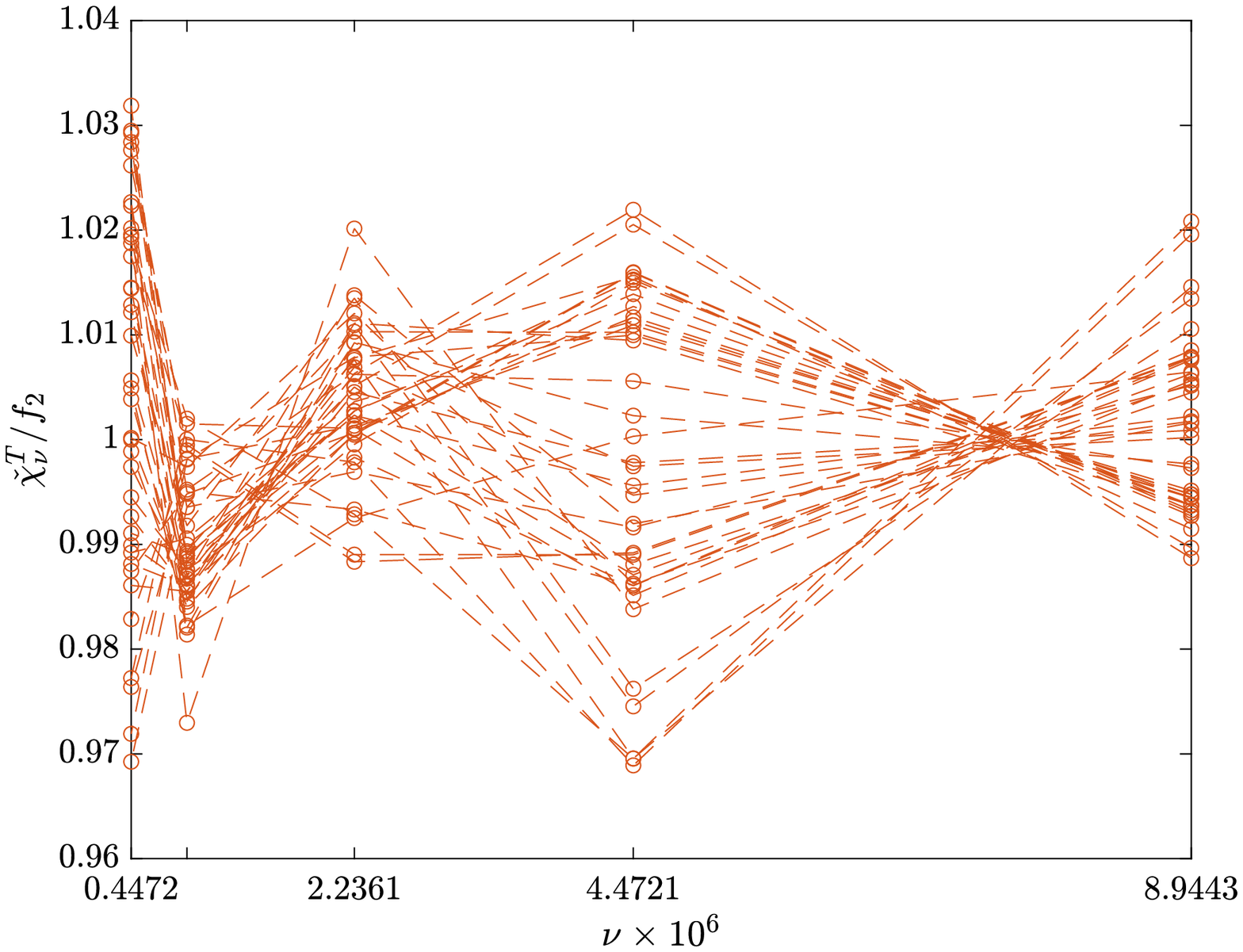}}}
\mbox{\subfigure[]
  {\includegraphics[width=0.5\textwidth]{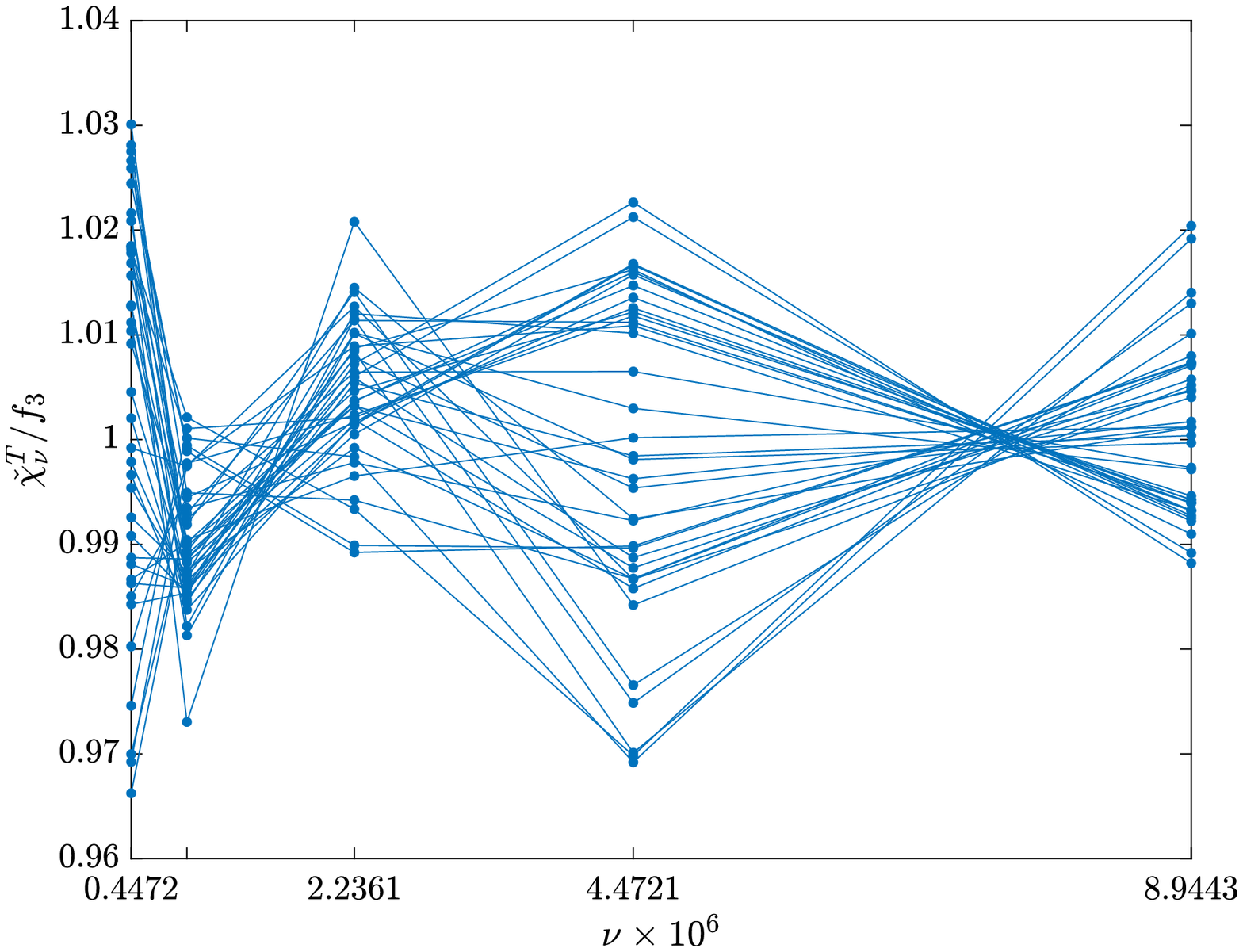}}\qquad
  \subfigure[]
  {\includegraphics[width=0.5\textwidth]{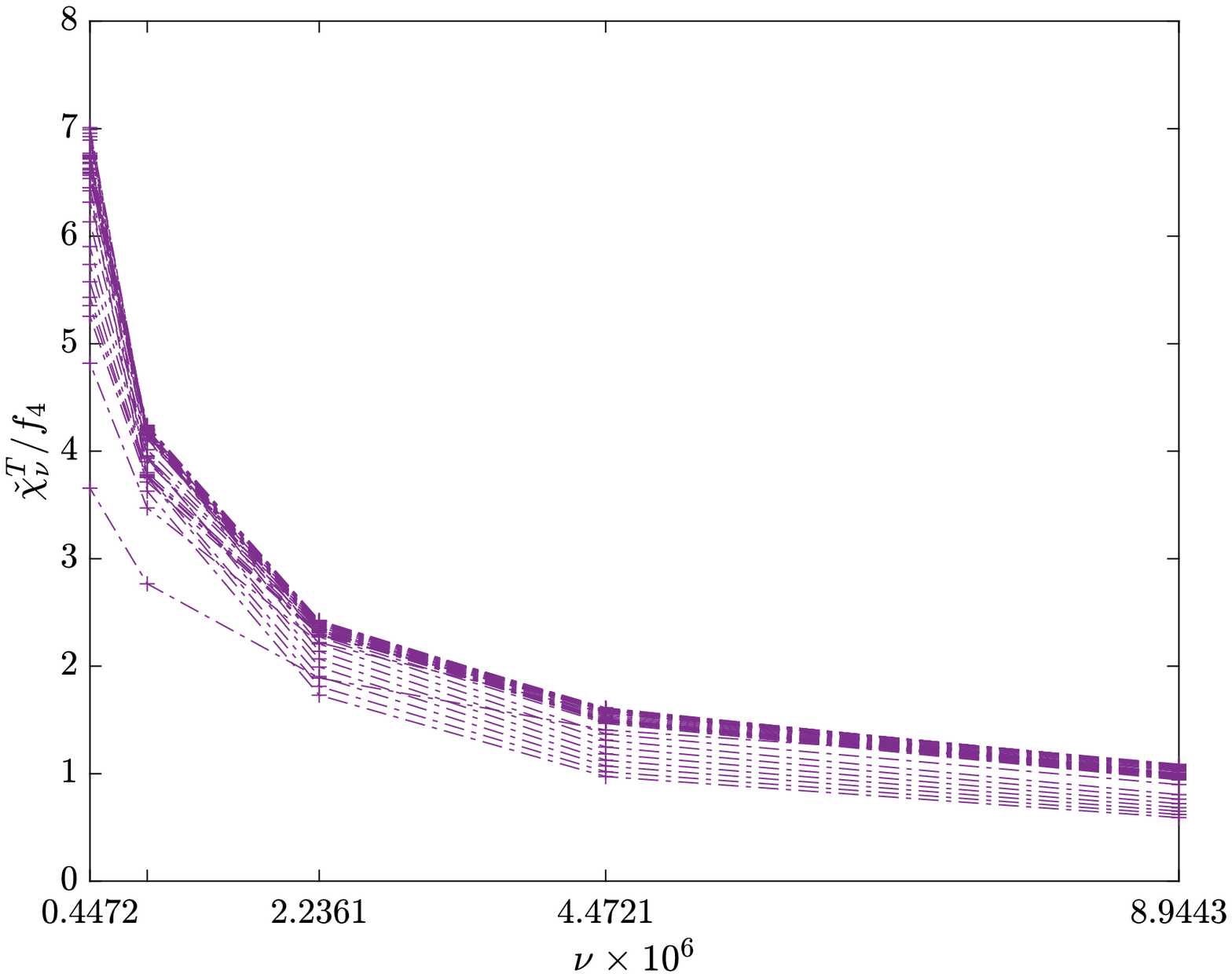}}}
\caption{Dependence of (a) $\chinck / f_1(\nu)$, (b)
  $\chinck / f_2(\nu)$, (c) $\chinck / f_3(\nu)$ and (d)
  $\chinck / f_4(\nu)$, with optimal constants {$\tC = \tC(T)$} and
  exponents $\talpha = \talpha(T)$, on the viscosity $\nu$ for
  different $T$.}
  \label{fig:ff}
\end{figure}

In order to assess now well the different ansatz functions
\eqref{eq:Tranf}--\eqref{eq:JYf} capture the dependence of the maximum
enstrophy dissipation $\chinck$ on $\nu$, cf.~Figure \ref{fig:chiT}b,
we define the ratios $\chinck / f_i(\nu)$, $i=1,2,3,4$, and plot them
as functions of $\nu$ for different $T$ in Figures \ref{fig:ff}a--d
using the values of {$\tC = \tC(T)$} and $\talpha = \talpha(T)$
determined as above. Thus, if $\chinck / f_i(\nu)$ is close to unity
over the entire range of $\nu$, this signals that the ansatz function
$f_i(\nu)$ accurately captures the dependence of $\chinck$ on $\nu$
for the given value of $T$.  We see that this is what indeed happens
for $f_2(\nu)$ and $f_3(\nu)$ for most values of $T$, cf.~Figures
\ref{fig:ff}b,c. On the other hand, we note that relations $f_1(\nu)$
and $f_4(\nu)$, respectively, overestimate and underestimate the
actual dependence of $\chinck$ on $\nu$, cf.~Figures
\ref{fig:ff}a,d. This observation is consistent with the fact that
\eqref{eq:Tranf} represents estimate \eqref{eq:Tran}, which is more
conservative than bounds \eqref{eq:chiCDE}--\eqref{eq:chiS}, and
\eqref{eq:JYf} has the form of the lower bound \eqref{eq:chiJY}.

\begin{figure}\centering
    \includegraphics[scale=0.5]{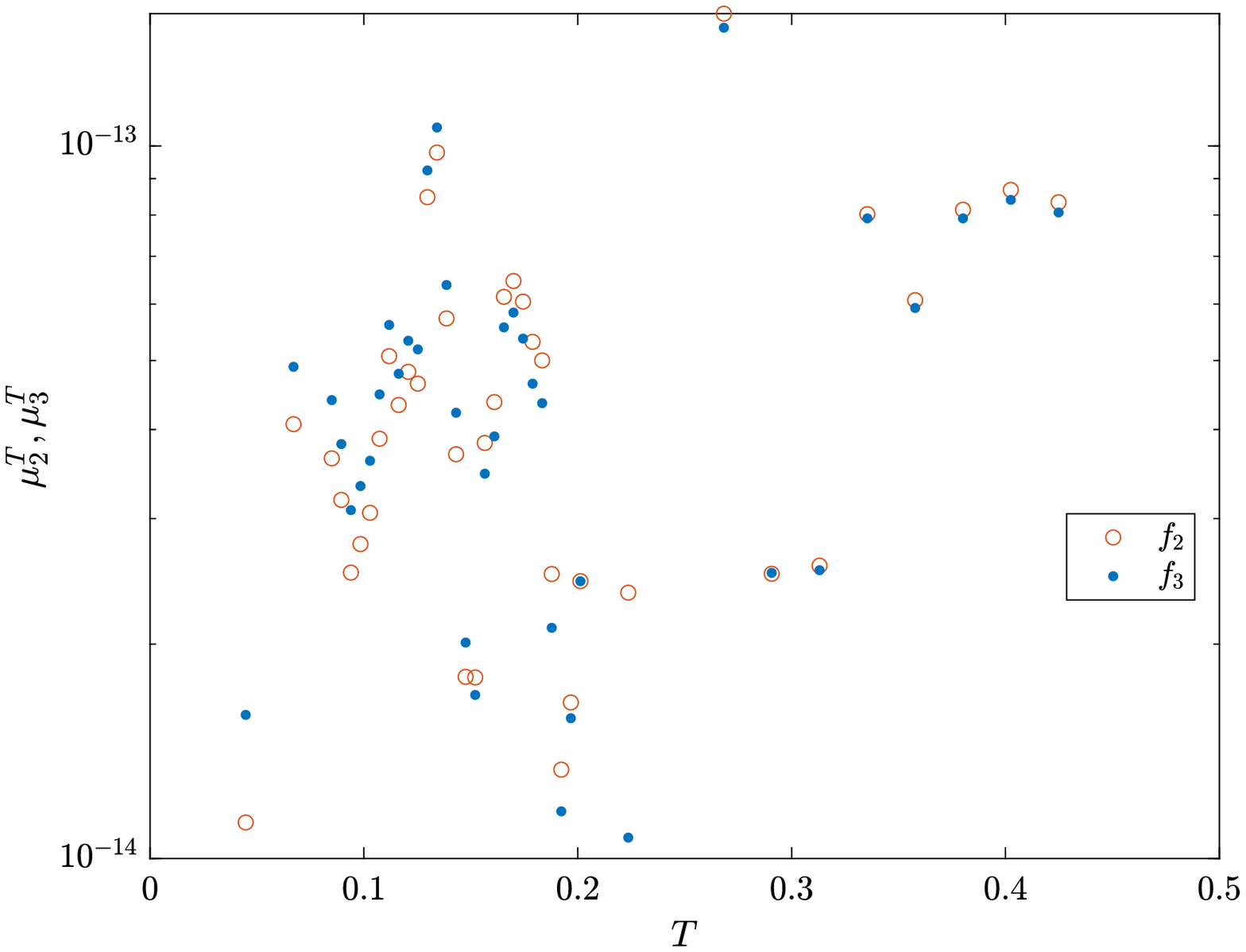}
    \caption{Dependence of the mean-square errors
      {$\mu_i^T(\tC(T))$,} $i=2,3$, cf.~\cref{eq:MSE},
      corresponding to the fits of ansatz functions (red circles)
      $f_2(\nu)$ and (blue dots) $f_3(\nu)$ to $\chinck$ for different
      $T$.}
  \label{fig:MSE}
\end{figure}

\begin{figure}\centering
    \includegraphics[scale=0.5]{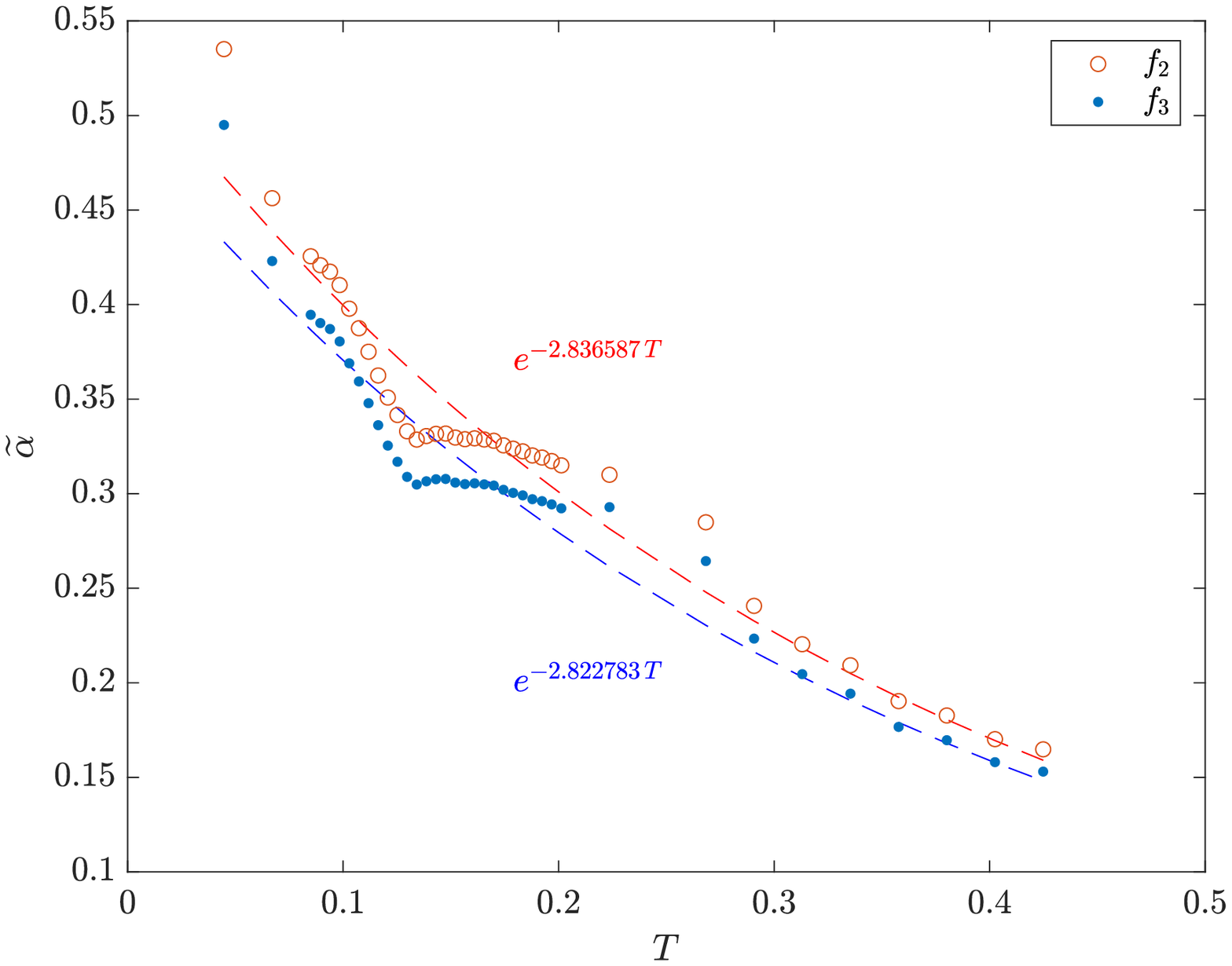}
    \caption{Dependence of the optimal exponents
      $\talpha = \talpha(T)$ in the ansatz functions (red circles)
      $f_2(\nu)$ and (blue dots) $f_3(\nu)$ on the length $T$ of the
      time window.  The dashed lines represent exponential fits,
      {in the forms indicated,} to the values of
      $\talpha = \talpha(T)$ for the ansatz function (red) $f_2(\nu)$
      and (blue) $f_3(\nu)$.}
 \label{fig:talpha}
\end{figure}

Hereafter we will focus on the fits given in terms of ansatz functions
\eqref{eq:CCSf}--\eqref{eq:Sf}. In order to decide which of these
relations more accurately represents the dependence of $\chinck$ on
$\nu$, in Figure \ref{fig:MSE} we show the corresponding mean-square
errors \eqref{eq:MSE} as functions of $T$. What this figure reveals is
that relation $f_2(\nu)$ generally leads to smallers errors for
shorter time windows (with $T \lessapprox 0.147$), whereas relation
$f_3(\nu)$ tends to better predict the dependence of $\chinck$ on
$\nu$ for longer time windows.  Finally, the optimal exponents
$\talpha = \talpha(T)$ determined for ansatz functions
\eqref{eq:CCSf}--\eqref{eq:Sf} are shown in Figure \ref{fig:talpha}
where an overall decreasing trend with $T$ is evident. As regards the
``dip'' occurring for $0.0894 \lessapprox T \lessapprox 0.1342$, we
speculate that it may be the result of some branches not being
captured in Figure \ref{fig:chiT}a. We note that, remarkably, the
dependence of the exponent $\talpha$ on $T$ reveals an approximately
exponential form consistent with the structure of the upper bounds in
\eqref{eq:CCSf}--\eqref{eq:Sf}. Moreover, the limit
$\lim_{T \rightarrow 0} \talpha(T)$ is also quantitatively consistent
with predictions of estimare \eqref{eq:Sf}.

\section{Summary and Conclusions}
\label{sec:conclusions}

In this study we provide a quantitative characterization of the
behaviour of the enstrophy dissipation in 2D Navier-Stokes flows in
the limit of vanishing viscosity. Unlike the case of Burgers flows in
1D and Navier-Stokes flows in 3D where the energy anomaly is well
documented, 2D Navier-Stokes flows are known not to exhibit anomalous
behavior of enstrophy dissipation. As discussed in Introduction, the
vanishing of enstrophy dissipation in the inviscid limit is subject to
various estimates, some ad-hoc and some rigorous, providing lower and
upper bounds on this quantity as viscosity vanishes. In our
investigation we have probed the sharpness of these estimates by
constructing families of Navier-Stokes flows designed to locally
maximize the enstrophy dissipation subject to certain
constraints. This was done by solving Problem \ref{pb:maxchi} where
locally optimal initial data $\phichk$ with fixed palinstrophy $\P_0$
was found such that the corresponding flow with the given viscosity
$\nu$ maximizes the enstrophy dissipation $\chin$ over the time window
$[0,T]$.  Problem \ref{pb:maxchi} was solved numerically using a
state-of-the-art adjoint-based gradient ascent method described in
Section \ref{sec:optimization}.  This optimization problem is
nonconvex and we have found six distinct branches of local maximizers,
each associated with a different mechanism for palinstrophy
amplification, cf.~Table \ref{tab:branches}. As is evident from
\href{https://youtu.be/4I_obQAgxUY}{Movie 1}, while in all cases
palinstrophy amplification involves stretching of thin vorticity
filaments, there are multiple ways to design flows maximizing this
process on a periodic domain $\Omega$ and which of these different
mechanisms produces the largest enstrophy dissipation depends on the
value of viscosity $\nu$ and the length $T$ of the time window,
cf.~Figure \ref{fig:chiT}a.

Branches of local maximiers found by solving Problem \ref{pb:maxchi}
for different values of $\nu$ and $T$ reveal how the extreme behaviour
of the enstrophy dissipation they realize compares with the available
estimates on this process discussed in Introduction. We conclude that
the dependence of the maximum enstrophy dissipation $\chinck$ in the
extreme flows we found on $\nu$ with fixed $T$ is quantitatively
consistent with the upper bound in estimate \eqref{eq:chiCCS},
cf.~Figure \ref{fig:ff}b, which is the sharpest estimate available to
date. Remarkably, the exponential dependence of the exponent in this
upper bound on $T$ is also quantitatively consistent with our results,
cf.~Figure \ref{fig:talpha} (we attribute the deviation from the
exponential decrease evident around $T \approx 0.1342$ in this figure
to the likely possibility that, despite our efforts, not all branches
of maximizing solutions have been found).

As regards estimate \eqref{eq:chiCCS}, we note that it depends on the
quantity $\| \phichk \|_{L^\infty(\Omega)}$ (via the constant
$M$). Since our optimal initial conditions are sought in the space
$H^1(\Omega)$, we do not have an a priori control over this quantity,
however, in our computations we did not find any evidence for
$\| \phichk \|_{L^\infty(\Omega)}$ to attain large values. Thus, these
caveats notwithstanding, we conclude that estimate \eqref{eq:chiCCS}
is sharp and does not offer any room for improvement, other than
perhaps a logarithmic correction analogous to the one appearing in
\eqref{eq:chiS}. Relation \eqref{eq:chiS} was found to describe the
dependence of the maximum enstrophy dissipation $\chinck$ on viscosity
in the limit $\nu \rightarrow 0$ with similar accuracy to estimate
\eqref{eq:chiCCS}. However, we {reiterate} that, as discussed in
Introduction, \eqref{eq:chiS} does not represent a rigorous upper
bound on the enstrophy dissipation. Improving this estimate, so that
the $\dot{H}^{-1}(\Omega)$ norm on the left-hand side in
\eqref{eq:chiS} is strengthened to $L^2(\Omega)$, appears to be an
open question in mathematical analysis.

Among other open problems, it would be interesting to better
understand the bifurcation structure of the different optimal solution
branches shown in Figure \ref{fig:chiT}a. Another open question is
what new insights about the problem considered here could be deduced
based on the kinetic theory, i.e., by considering an optimization
problem analogous to Problem \ref{pb:maxchi} in the context of the
Boltzmann equation or some of its variants. Some efforts in this
direction are already underway. Finally, there is the question
about what can be said about the energy dissipation anomaly in 3D
Navier-Stokes flows using the approach developed in the present study.

\section*{Acknowledgments}

This work is dedicated to the memory of the late Charlie Doering, our
dear friend and collaborator, who served as an inspiration to so many
of us. It is related to our long-term research program influenced by
Charlie's interest in questions concerning saturation of rigorous
bounds.

The authors thank Roman Shvydkoy for interesting discussions
and for bringing a number of relevant references to their attention.

The first two authors acknowledge the support through an NSERC
(Canada) Discovery Grant. The second author would like to thank the
Isaac Newton Institute for Mathematical Sciences for support and
hospitality during the programme ``Mathematical aspects of turbulence:
where do we stand?'' where this work was finalized. This work was
supported by EPSRC grant number EP/R014604/1.  The research of the
third author was partly supported by the JSPS Grants-in-Aid for
Scientific Research 20H01819.  The authors also thank the Japan
Society for the Promotion of Science for awarding the first author a
research fellowship to conduct this work.  Computational resources
were provided by Compute Canada under its Resource Allocation
Competition.



\end{document}